\documentclass[prb,aps,amssymb,twocolumn,superscriptaddress,notitlepage]{revtex4-2}
\usepackage{amsmath}
\usepackage{amssymb}
\usepackage{amsthm}
\usepackage{amsfonts}
\usepackage{listings}
\usepackage{physics}
\usepackage{enumerate}
\usepackage{latexsym}
\usepackage{psfrag}
\usepackage{bm}
\usepackage{graphicx}
\usepackage[caption=false]{subfig}
\usepackage{blkarray}
\usepackage{array}
\usepackage{color}
\usepackage[normalem]{ulem}
\usepackage{hyperref}
\hypersetup{
     colorlinks = true,
     linkcolor = magenta,
     citecolor = magenta,
     breaklinks= true
}

\def\beq{\begin{equation}}
\def\eeq{\end{equation}}
\def\bal{\begin{aligned}}
\def\eal{\end{aligned}}

\newcommand{\com}[2]{\left[#1,#2\right]}

\begin{document}

\title{Spectral Density and Sum Rules for Second-Order Response Functions}

\author{Barry Bradlyn}
\email{bbradlyn@illinois.edu}
\affiliation{Department of Physics, University of Illinois at Urbana-Champaign, Urbana IL 61801, USA}
\affiliation{Anthony J. 
Leggett Institute for Condensed Matter Theory, University of Illinois at Urbana-Champaign, Urbana IL 61801, USA}
\author{Peter Abbamonte}
\affiliation{Department of Physics, University of Illinois at Urbana-Champaign, Urbana IL 61801, USA}
\affiliation{Materials Research Laboratory, University of Illinois at Urbana-Champaign, Urbana, IL 61801, USA}
\date{\today}

\begin{abstract}
Sum rules for linear response functions give powerful and experimentally-relevant relations between frequency moments of response functions and ground state properties. 
In particular, renewed interest has been drawn to optical conductivity and density-density sum rules and their connection to quantum geometry in topological materials. 
At the same time, recent work has also illustrated the connection between quantum geometry and second-order nonlinear response functions in quantum materials, motivating the search for exact sum rules for second-order response that can provide experimental probes and theoretical constraints for geometry and topology in these systems. 
Here we begin to address these questions by developing a general formalism for deriving sum rules for second-order response functions. 
Using generalized Kramers-Kronig relations, we show that the second-order Kubo formula can be expressed in terms of a spectral density that is a sum of Dirac delta functions in frequency. 
We show that moments of the spectral density can be expressed in terms of averages of equal-time commutators, yielding a family of generalized sum rules; furthermore, these sum rules constrain the large-frequency asymptotic behavior of the second harmonic generation rate. 
We apply our formalism to study generalized $f$-sum rules for the second-order density-density response function and the longitudinal nonlinear conductivity. 
We show that for noninteracting electrons in solids, the generalized $f$-sum rule can be written entirely in terms of matrix elements of the Bloch Hamiltonian.  
Finally, we derive a family of sum rules for rectification response, determining the large-frequency asymptotic behavior of the time-independent response to a harmonic perturbation.
\end{abstract}

\maketitle

\section{Introduction}\label{sec:intro}
Sum rules for response functions represent some of the few exact results in many-body physics~\cite{kadanoff1963hydrodynamic,kubo1957statistical}.
They are of profound fundamental importance, and are critical for interpreting spectroscopy experiments where they are used to calibrate the absolute scale of measurements \cite{Dressel2002,PaulMartin1968,Boothroyd2020}.
The best known sum rule, known as the $f$-sum rule, relates the total electron density in a solid to the absorption probability of light, placing rigorous constraints on optical properties of materials.
Furthermore, the $f$-sum rule is of practical utility to experimentalists when it comes to extrapolating the frequency dependence of measured response functions~\cite{tanuma1993use}.
The related conductivity sum rule analogously determines the frequency of plasma oscillations in terms of the frequency-integrated AC susceptibility of a solid.
Via Kramers-Kronig relations, sum rules also tell us about the large-frequency behavior of response functions~\cite{kadanoff1963hydrodynamic,forster1975hydrodynamic,bradlyn2012kubo}. 
These linear sum rules provide crucial self-consistency checks on experimental measurements of optical constants, as well as allow for the comparison of measurements using different techniques and in different frequency ranges, enabling the compilation of composite datasets of optical properties~\cite{smith1997dispersion}.

Recently, attention has been paid to sum rules derived for more general moments of the conductivity tensor. 
In particular, Ref.~\cite{souza2000polarization} derived a sum rule relating the first negative moment of the optical (AC) conductivity tensor to polarization fluctuations in the ground state of a many body system. 
For noninteracting insulators, the polarization fluctuations can be expressed in terms of the quantum geometry of the occupied electronic states via the Fubini-Study metric, which has proved important in determining the superfluid weight in flat band systems~\cite{torma2022superconductivity}. 
More generally, for interacting systems the polarization fluctuation sum rule can be expressed in terms of the ``quantum weight'' that generalizes the Fubini-Study metric to interacting systems~\cite{souza2000polarization,onishi2024quantum}. 
The connection between sum rules and quantum geometry has brought renewed interest to the study of linear response functions.

However, outside of specialized applications to nonlinear optics~\cite{scandolo1995kramerskronig,bassani1991dispersion,bassani1992sum,bassani1998general,jha1984nonlinear}, the main focus of study has been sum rules for linear response functions.
Notable exceptions are Refs.~\cite{watanabe2020generalized,watanabe2020general} which considered a family of nonlinear conductivity sum rules derived from the instantaneous current response to an applied (spatially uniform) electric field, as well as Ref.~\cite{matsyshyn2019nonlinear} which derives a sum rule for the nonlinear Hall conductivity. 
However, the resurgence of interest in nonlinear---and in particular second-order---response functions in the study of topological materials~\cite{dejuan2017quantized,sodemann2015quantum,ahn2022riemannian,morimoto2016topological,morimoto2016semiclassical,rees2020helicitydependent,ahn2020lowfrequency,flicker2018chiral,kaushik2021magnetic,mckay2021optical} motivates the study of nonlinear sum rules more systematically. 
For instance, Ref.~\cite{ahn2022riemannian} has shown that the nonlinear conductivity of quantum systems is connected to the Riemannian geometry of quantum states. 
Sum rules for nonlinear response functions could thus place important constraints on, and provide experimental probes for, quantum geometry in insulators. 
They would also enable experimenters to properly calibrate nonlinear spectroscopy measurements in the same manner customary for linear spectroscopies.

In this work, we will take the first steps towards addressing these questions by studying sum rules for second-order response functions. 
We will ultimately derive a set of sum rules for second harmonic generation response of a system to general perturbation. 
First, in Sec.~\ref{sec:2ndorder} we will review the Kubo formula for second-order response functions. 
We will introduce a representation for the response function in terms of a spectral density, generalizing the Kramers-Kronig relations of linear response theory. 
Next, in Sec.~\ref{sec:sumrule} we will show that frequency moments of the spectral density determine the large-frequency behavior of the second-order response function. 
Furthermore, we will show that these frequency moments can be expressed in terms of averages of equal-time commutators, thus furnishing a set of second-order sum rules. 
We will show that by focusing on second harmonic generation response, we arrive at a sum rule relevant to experiments. 
In Sec.~\ref{sec:density} we will specialize to consider the second-order density response function $\chi_{\rho\rho}^{(2)}$, which quantifies the response of the local charge density $\rho(\mathbf{r})$ to a scalar potential perturbation at second order. 
Applying our sum rules to $\chi_{\rho\rho}^{(2)}$ allows us to derive a generalization of the $f$-sum rule relevant to nonlinear spectroscopy. 
This sum rule could enable determination of absolute cross sections for nonlinear optical processes such as second harmonic generation, parametric down conversion, or optical rectification.
As an example of our formalism, in Sec.~\ref{sec:freefermions} we evaluate our nonlinear $f$-sum rule for a system of free fermions in a periodic potential. 
By exploiting charge conservation, we show in Sec.~\ref{sec:cond} that our nonlinear $f$-sum rule also implies a sum rule for the longitudinal component of the nonlinear conductivity tensor, applicable to optical experiments. 
Finally, in Sec.~\ref{sec:rect} we turn our attention to second-order rectification, the time-independent response to a harmonic perturbation. 
We show that for operators that can be written as time derivatives (such as the uniform current, which is the time derivative of the position operator), the large frequency asymptotic behavior of the rectification response is determined by a family of sum rules 
We conclude in Sec.~\ref{sec:outlook} with a discussion of potential experimental and theoretical applications. 

For notational convenience, we will work in units where $\hbar=c=e=1$. 
Additionally, for integrals over the domain $(-\infty,\infty)$ we will leave the limits of integration implicit.

\section{Second-Order Response and the Spectral Density}\label{sec:2ndorder}

Our setting for studying nonlinear response is a system with unperturbed Hamiltonian $H_0$, coupled to an external field. 
We model the coupling to the external field by the perturbing Hamiltonian
\begin{equation}
H_1(t)=e^{\epsilon t}f(t)B,
\end{equation} 
where $f(t)$ is the amplitude of the externally applied field, $B$ is the operator to which the field couples, and $\epsilon$ is a small positive infinitesimal that we will take to zero at the end of any calculation (it ensures that the perturbation goes to zero infinitely far in the past). 
We will be interested in how the average $\langle A\rangle(t)$ of some observable $A$ depends on the perturbation $f(t)$. 
We can expand in powers of $f(t)$ to write
\begin{equation}
\langle A\rangle(t) = \langle A\rangle_0 + \delta\langle
A\rangle(t) + \delta^2\langle
A\rangle(t)+\dots,
\end{equation}
where $\langle A\rangle_0$ is the unperturbed average of $A$, and $\delta^n\langle
A\rangle(t)$ is the correction to the average of $A$ to $n$-th order in the perturbation $f$. 
The linear response $\delta\langle
A\rangle(t)$ can be expressed using the standard Kubo formula, which we re-derive for completeness in Appendix~\ref{sec:derivations}. 
Going to one higher order in perturbation theory, we can write the second-order response $\delta^2\langle
A\rangle(t)$ in terms of a nonlinear response function
$\chi^{(2)}_{AB}$, defined via the constitutive relation
\begin{align}\label{eq:chi2def}
\delta^2\langle A\rangle(t)
&=\int dt'dt''\chi^{(2)}_{AB}(t-t',t-t'')f(t')f(t''),
\end{align}
As derived in Appendix~\ref{sec:derivations}, we can take the Fourier transform of Eq.~\eqref{eq:chi2def} to find
\begin{widetext}
\begin{equation}\label{eq:chi2deffourier}
\delta^2\langle A\rangle(\omega)=\int  d\omega_1d\omega_2 \delta(\omega-\omega_1-\omega_2)\chi^{(2)}_{AB}(\omega_1,\omega_2)f(\omega_1)f(\omega_2),
\end{equation}
where 
\begin{equation}\label{eq:chi2finalmain}
\chi^{(2)}_{AB}(\omega_1,\omega_2)=-\int_{0}^{\infty}du\int_{0}^{\infty}dv \langle\com{\com{A(u+v)}{B(v)}}{B(0)}\rangle_0e^{i\omega_{12}^+ u}e^{i\omega_2^+ v},
\end{equation}
\end{widetext}
and where we have introduced the shorthand
\begin{align}\label{eq:w12}
\omega_{1}^+ &= \omega_1+i\epsilon, \nonumber\\
\omega_{2}^+ &= \omega_2+i\epsilon, \nonumber\\
\omega_{12}^+&=\omega_1^++\omega_2^+.
\end{align}

Eq.~\eqref{eq:chi2finalmain} is the second-order generalization of the Kubo formula. 
Equivalent expressions appeared in the literature as early as Ref.~\cite{kubo1957statistical}, and equivalent formulas have been widely applied more recently to the study of electromagnetic response functions~\cite{ahn2022riemannian,watanabe2020generalized,watanabe2020general,sipe2000secondorder,parker2019diagrammatic,morimoto2016topological,jha1984nonlinear}. 
Here, we rewrite the nonlinear Kubo formula more generally, and in terms of time variables $u$ and $v$ that makes causality manifest: note that in Eq.~\eqref{eq:chi2finalmain} $A(u+v)$ is evaluated at a later time than $B(v)$, which itself is evaluated at a later time than $B(0)$. 

Note, importantly, that the function $\chi^{(2)}_{AB}(\omega_1,\omega_2)$ is not itself a direct experimental observable. 
This is because the perturbing fields on the right hand side of Eq.~\eqref{eq:chi2deffourier} are symmetric under the interchange of integration variables $\omega_1\leftrightarrow\omega_2$. 
The response function computed from an experimental measurement of $\delta^2\langle A\rangle(\omega)$ is thus not $\chi^{(2)}_{AB}(\omega_1,\omega_2)$ but instead
\begin{equation}\label{eq:chi2physical}
\bar{\chi}^{(2)}_{AB}(\omega_1,\omega_2)\equiv\frac{1}{2}\left(\chi^{(2)}_{AB}(\omega_1,\omega_2)+\chi^{(2)}_{AB}(\omega_2,\omega_1)\right).
\end{equation}
With this in mind, we will use Eq.~\eqref{eq:chi2finalmain} for deriving general properties of the response function, and will explicitly symmetrize using Eq.~\eqref{eq:chi2physical} before presenting physically-meaningful results. 

As we discuss in Appendix~\ref{sec:derivations}, causality implies that both $\chi^{(2)}_{AB}(\omega_1,\omega_2)$ and $\bar{\chi}^{(2)}_{AB}(\omega_1,\omega_2)$ are analytic functions in the upper half $\omega_1$ and $\omega_2$ planes. 
As such, we can express $\chi^{(2)}_{AB}(\omega_1,\omega_2)$ as an integral over a spectral density. 
By using the Fourier representation of the step function  [see Eq.~\eqref{eq:thetafnfourier}], we can write
\begin{equation}\label{eq:chi2specdensityrep}
\chi^{(2)}_{AB}(\omega_1,\omega_2) = \frac{1}{\pi^2}\int  d\alpha d\beta \frac{\chi^{(2)''}_{AB}(\alpha,\beta)}{(\alpha-\omega_{12}^+)(\beta-\omega_2^+)}.
\end{equation}
%\begin{widetext}
The spectral density $\chi^{(2)''}_{AB}(\alpha,\beta)$ is given as the Fourier transformed correlation function
\begin{equation}\label{eq:chi2specdensity}
\chi^{(2)''}_{AB}(\alpha,\beta) = \frac{1}{4}\int  du dv \langle\com{\com{A(u+v)}{B(v)}}{B(0)}\rangle_0e^{i(\alpha u + \beta v)}.
\end{equation}
Eq.~\eqref{eq:chi2specdensity} is the first main result of this work. 
Note that $\chi^{(2)''}_{AB}(\alpha,\beta)$ is the Fourier transform of the correlation function appearing in the time-domain expression Eq.~\eqref{eq:chi2timedomain} of the nonlinear response functions. 
Thus, the spectral density also has the interpretation of an instantaneous response function, in analogy with the spectral density for linear response functions~\cite{watanabe2020generalized,verma2024instantaneous}. 
Additionally, similar to the spectral density for linear response functions [see Eq.~\eqref{eq:specdensity1lehmann} in Appendix~\ref{sec:linreview}], we can write $\chi^{(2)''}_{AB}(\alpha,\beta)$ in the Lehmann representation as a weighted sum of Dirac delta functions. 
We consider a ground state stationary density matrix of the form
\begin{equation}\label{eq:gndstate}
\rho_0 = \sum_n p_n\outerproduct{n}{n},
\end{equation}
where $\ket{n}$ is a complete set of eigenstates of the unperturbed Hamiltonian $H_0$ with energy $E_n$, each occupied with probability $p_n$. 
Inserting complete sets of states in Eq.~\eqref{eq:chi2specdensity} and evaluating the average we find
\begin{widetext}
\begin{align}\label{eq:specdensitydelta}
\chi^{(2)''}_{AB}(\alpha,\beta) = \pi^2 \sum_{nm\ell}&\Big[\mel{\ell}{A}{n}\mel{n}{B}{m}\mel{m}{B}{\ell}\delta(\alpha+E_\ell-E_n)\delta(\beta+E_\ell-E_m)(p_\ell-p_m)\Big. \nonumber \\
&+\Big.\mel{\ell}{B}{n}\mel{n}{B}{m}\mel{m}{A}{\ell}\delta(\alpha+E_m-E_\ell)\delta(\beta+E_n-E_\ell)(p_\ell-p_n)\Big].
\end{align}
\end{widetext}

The spectral representation Eq.~\eqref{eq:chi2specdensityrep} further allows us to derive generalized Kramers-Kronig relations for the second-order response function. 
Using the Plemelj formula [see Eq.~\eqref{eq:plemelj}] to rewrite the denominators in Eq.~\eqref{eq:chi2specdensityrep} and exploiting the analyticity of $\chi^{(2)}_{AB}(\omega_1,\omega_2)$ in the upper half $\omega_1$ and $\omega_2$ planes yields the decomposition
\begin{align}
\chi^{(2)}_{AB}(\omega_1,\omega_2) &=  \chi_{AB}^{(2),1}(\omega_{12},\omega_2) + i\chi_{AB}^{(2),2}(\omega_{12},\omega_2),
\end{align}
where we have defined 
\begin{align}
\chi_{AB}^{(2),1}(x,y)&=\frac{1}{\pi^2}\mathrm{P}\int  d\alpha d\beta \frac{\chi^{(2)''}_{AB}(\alpha,\beta)}{(\alpha-x)(\beta-y)} %\nonumber \\&
-\chi^{(2)''}_{AB}(x,y),\label{chi2decomp1}  \\
\chi_{AB}^{(2),2}(x,y)&=\frac{1}{\pi}\mathrm{P}\int  d\alpha \left[\frac{\chi^{(2)''}_{AB}(\alpha,y)}{\alpha-x} + \frac{\chi^{(2)''}_{AB}(x,\alpha)}{\alpha-y}\right], \label{eq:chi2decomp}
\end{align}
and used $\mathrm{P}$ to denote the Cauchy principal value. 
Symmetrizing Eqs.~\eqref{chi2decomp1} and \eqref{eq:chi2decomp} using Eq.~\eqref{eq:chi2physical}, we can write for the physical response function
\begin{equation}
\bar{\chi}^{(2)}_{AB}(\omega_1,\omega_2) = \bar{\chi}_{AB}^{(2),1}(\omega_{1},\omega_2) + i\bar{\chi}_{AB}^{(2),2}(\omega_{1},\omega_2)
\end{equation}
with
\begin{align}
\bar{\chi}_{AB}^{(2),1}(\omega_{1},\omega_2)&=\frac{1}{2}\left( \chi_{AB}^{(2),1}(\omega_{12},\omega_1)+ \chi_{AB}^{(2),1}(\omega_{12},\omega_2) \right), \\
\bar{\chi}_{AB}^{(2),2}(\omega_{1},\omega_2)&=\frac{1}{2}\left( \chi_{AB}^{(2),2}(\omega_{12},\omega_1)+ \chi_{AB}^{(2),2}(\omega_{12},\omega_2) \right).
\end{align}

Unfortunately, unlike in the linear Kramers-Kronig relation, neither $\chi_{AB}^{(2),1}(\omega_{12},\omega_2)$ nor $\chi_{AB}^{(2),2}(\omega_{12},\omega_2)$ are proportional to the spectral density $\chi^{(2)''}_{AB}(\omega_{12},\omega_2)$, meaning that direct measurement of the nonlinear spectral density $\chi^{(2)''}_{AB}(\alpha,\beta)$ is not straightforward. 
Nevertheless, $\chi_{AB}^{(2),1}(\omega_{12},\omega_2)$ and $\chi_{AB}^{(2),2}(\omega_{12},\omega_2)$ still form a Hilbert transform pair, generalizing the linear Kramers-Kronig relations. 
In particular, 
\begin{align}
\chi_{AB}^{(2),1}(x,y)&=\frac{1}{\pi}\mathrm{P}\int d\alpha \frac{\chi_{AB}^{(2),2}(\alpha,y)}{\alpha-x} \nonumber \\
&= \frac{1}{\pi}\mathrm{P}\int d\alpha \frac{\chi_{AB}^{(2),2}(x,\alpha)}{\alpha-y}, \\
\chi_{AB}^{(2),2}(x,y)&=-\frac{1}{\pi}\mathrm{P}\int d\alpha \frac{\chi_{AB}^{(2),1}(\alpha,y)}{\alpha-x} \nonumber \\
&= -\frac{1}{\pi}\mathrm{P}\int d\alpha \frac{\chi_{AB}^{(2),1}(x,\alpha)}{\alpha-y}.
\end{align}
Similar generalized Kramers-Kronig relations have been previously obtained in, e.g. 
Refs.~\cite{scandolo1995kramerskronig,bassani1991dispersion,bassani1992sum,bassani1998general}, although the connection with the spectral density was not emphasized in those works. 
Note that, because these relationships do not directly involve the absorptive and reactive parts of the response, they cannot be applied to nonlinear spectroscopic data in the same, straightforward way that the usual Kramers-Kronig relations can be applied to linear spectroscopies. 
However, they could be implemented in an iterative, self-consistent scheme to accomplish the same purpose.

Although the second-order spectral density Eq.~\eqref{eq:chi2specdensityrep} is not directly measurable through absorption, it can still be probed experimentally through the high-frequency behavior of the response function. 
In particular, we will see now that the high-frequency asymptotic decay of $\bar\chi^{(2)}(\omega_1,\omega_2)$ is determined by frequency moments of the spectral density $\chi^{(2)''}(\alpha,\beta)$.

\section{Asymptotics and Sum Rules for Second Harmonic Generation}\label{sec:sumrule}

One main utility of the spectral representation Eq.~\eqref{eq:chi2specdensityrep} is that it allows us to derive sum rules for the large frequency asymptotic behavior of $\bar{\chi}^{(2)}(\omega_1,\omega_2)$. 
Formally, we can expand Eq.~\eqref{eq:chi2specdensityrep} in the limit of large $\omega_1$ and $\omega_2$ of the same sign and, after symmetrizing under the exchange of $\omega_1\leftrightarrow\omega_2$ [as per Eq.~\eqref{eq:chi2physical}] we obtain the asymptotic expansion
\begin{widetext}
\begin{align}
\bar{\chi}^{(2)}(\omega_1\rightarrow\infty,\omega_2\rightarrow\infty)\sim\frac{1}{2\pi^2}\sum_{n,m=0}^{\infty}\frac{1}{(\omega_1+\omega_2)^{n+1}}\left(\frac{1}{\omega_1^{m+1}}+\frac{1}{\omega_2^{m+1}}\right)\int  d\alpha d\beta \left[\alpha^n\beta^m \chi^{(2)''}_{AB}(\alpha,\beta)\right].\label{eq:chi2asymptotic}
\end{align}
\end{widetext}
Since this asymptotic expansion only holds for both $\omega_1$ and $\omega_2$ large, its physical significance is somewhat obscure. 
However, we can apply Eq.~\eqref{eq:chi2asymptotic} to the case of second harmonic generation $\omega_1=\omega_2=\omega$. 
We define the second harmonic generation response function
\begin{equation}
\chi_{AB}^{\mathrm{SHG}}(\omega) = \bar{\chi}^{(2)}(\omega,\omega) = {\chi}^{(2)}(\omega,\omega)
\end{equation}
which determines the leading order contribution to the $2\omega$ Fourier component of $\langle A\rangle$ in response to a harmonic drive $f(\omega)$ at frequency $\omega$ via
\begin{equation}
\langle A\rangle(2\omega) = \chi_{AB}^{\mathrm{SHG}}(\omega)[f(\omega)]^2 + \mathcal{O}(f^4).
\end{equation}
Setting $\omega_1=\omega_2=\omega$ in the asymptotic expansion Eq.~\eqref{eq:chi2asymptotic}, we find that the second harmonic generation response at large frequency is given by
\begin{align}\label{eq:shgasymptotic}
\bar{\chi}^{\mathrm{SHG}}_{AB}(\omega\rightarrow\infty)%&\sim \frac{1}{2\pi^2\omega^2}\sum_{N=0}^{\infty}\sum_{m=0}^N \frac{1}{\omega^N}\int d\alpha\int d\beta \left(\frac{\alpha}{2}\right)^{N-m}\beta^m\chi^{(2)''}_{AB}(\alpha\beta) \\
&\sim\sum_{N=0}^{\infty} \frac{\nu^{(N)}_{AB}}{2\omega^{N+2}},
\end{align}
where we have defined the weighted moments $\nu^{(N)}_{AB}$ as
\begin{equation}
\nu^{(N)}_{AB}=\frac{1}{\pi^2}\sum_{m=0}^N \int d\alpha d\beta \left(\frac{\alpha}{2}\right)^{N-m}\beta^m\chi^{(2)''}_{AB}(\alpha,\beta).
\end{equation}
We thus see that the moments $\nu^{(N)}_{AB}$ of the nonlinear spectral density $\chi^{(2)''}_{AB}(\alpha,\beta)$ determine the asymptotic decay of the second harmonic generation response at large frequency. 
Furthermore, since $\chi^{(2)''}_{AB}(\alpha,\beta)$ is given in Eq.~\eqref{eq:chi2specdensity} as the Fourier transform of a correlation function, we can express the moments $\nu^{(N)}_{AB}$ in terms of equal time correlation functions. 
For concreteness, we can evaluate the first three moments $N=0,1$ and $2$ explicitly. 
From Eqs.~\eqref{eq:shgasymptotic} and \eqref{eq:chi2specdensity} we find
\begin{align}\label{eq:shgnu0}
\nu^{(0)}_{AB} &= \frac{1}{\pi^2}\int d\alpha d\beta\chi^{(2)''}_{AB}(\alpha,\beta) \nonumber \\
%&=\frac{1}{4\pi^2}\int d\alpha\int d\beta\int du\int dv\langle\com{\com{A(u+v)}{B(v)}}{B(0)}\rangle_0e^{i\alpha u + i \beta v} \\
&=\langle\com{\com{A}{B}}{B}\rangle_0.
\end{align}
Similarly, for $\nu^{(1)}_{AB}$ we find
\begin{align}\label{eq:shgnu1}
\nu^{(1)}_{AB} &= \frac{1}{\pi^2}\int d\alpha d\beta\left(\frac{\alpha}{2}+\beta\right)\chi^{(2)''}_{AB}(\alpha,\beta)\nonumber \\
%&=\frac{1}{4\pi^2}\int d\alpha\int d\beta\int du\int dv\langle\com{\com{A(u+v)}{B(v)}}{B(0)}\rangle_0\left(\frac{\alpha}{2}+\beta\right)e^{i\alpha u + i \beta v} \\
&=\langle\com{i\partial_t\com{A}{B}}{B}\rangle_0 + \frac{1}{2}\langle\com{\com{i\partial_t A}{B}}{B}\rangle_0.
\end{align}
Finally, for $\nu^{(2)}_{AB}$ we have
\begin{align}\label{eq:shgnu2}
\nu^{(2)}_{AB} &= \frac{1}{\pi^2}\int d\alpha d\beta\left(\frac{\alpha^2}{4}+\beta^2+\frac{\alpha\beta}{2}\right)\chi^{(2)''}_{AB}(\alpha,\beta) \nonumber \\
&=-\langle\com{\partial^2_t\com{A}{B}}{B}\rangle_0 - \frac{1}{2}\langle\com{\partial_t\com{\partial_t A}{B}}{B}\rangle_0 \nonumber \\
&-\frac{1}{4}\langle\com{\com{\partial^2_t A}{B}}{B}\rangle_0.
\end{align}

Eqs.~\eqref{eq:shgnu0}--\eqref{eq:shgnu2} relate the frequency moments of the spectral density $\chi^{(2)''}_{AB}$ to ground state averages of commutators of $A$, $B$, and their time derivatives. 
They are nonlinear generalizations of the standard sum rules for the linear response coefficients introduced in Refs.~\cite{kubo1957statistical,kadanoff1963hydrodynamic}, and reviewed in Appendix~\ref{sec:linreview}. 
Comparing with the time-domain expression for the nonlinear response function [i.e., Eq.~\eqref{eq:chi2timedomain} and the integrand in Eq.~\eqref{eq:chi2finalmain}], we see that the averages appearing in the sum rule correspond to a gradient expansion of the time-domain response near $u=v=0$, sometimes called the instantaneous response function~\cite{watanabe2020general,verma2024instantaneous}. 

We see that the sum rules constrain the large frequency behavior of the second harmonic generation response via Eq.~\eqref{eq:shgasymptotic}, which can be measured via nonlinear spectroscopic techniques. 
Furthermore, the sum rules and asymptotic expansions derived here make no assumptions on the unperturbed Hamiltonian other than that it conserves energy, and that averages are taken in a state described by a stationary density matrix. 
In particular, the asymptotic expansion Eq.~\eqref{eq:shgasymptotic} and the sum rules Eqs.~\eqref{eq:shgnu0}--\eqref{eq:shgnu2} (as well as their generalization to higher moments) hold for arbitrary interacting many-body systems. 
In the next section, we will apply this formalism to analyze the response of the electronic charge density $\rho_\mathbf{q}$ to the scalar electric potential $\phi_\mathbf{q}(t)$.

\section{Density response and the Second-order $f$-sum Rule}\label{sec:density}

We will now study the implications of our nonlinear sum rules Eqs.~\eqref{eq:shgnu0}--\eqref{eq:shgnu2} for second-order density response. 
We consider the response of the charge density operator
\begin{equation}
A=\rho_{\mathbf{q}}
\end{equation}
to a scalar electric potential perturbation
\begin{equation}
H_1=\sum_{\mathbf{q}}\phi_\mathbf{q}(t)\rho_{-\mathbf{q}},
\end{equation}
where $\phi_\mathbf{q}(t)$ is that (spatial) Fourier transform of the electric potential. 
Following our derivation of the nonlinear response function in Appendix~\ref{sec:derivations}, we can define the second order density-density response function $\chi^{(2)}_{\rho\rho}(\omega_1,\mathbf{q}_1,\omega_2,\mathbf{q}_2)$ via
\begin{widetext}
\begin{equation}
\delta^2\langle\rho_\mathbf{q}\rangle(\omega) = \sum_{\mathbf{q_1},\mathbf{q_2}}\int d\omega_1 d\omega_2\chi^{(2)}_{\rho\rho}(\omega_1,\mathbf{q}_1,\omega_2,\mathbf{q}_2)\delta(\mathbf{q}-\mathbf{q}_{12})\delta(\omega-\omega_{12})f_{\mathbf{q}_1}(\omega_1)f_{\mathbf{q}_2}(\omega_2),
\end{equation}
where the nonlinear Kubo formula gives
\begin{equation}\label{eq:chirho2kubo}
\chi^{(2)}_{\rho\rho}(\omega_1,\mathbf{q}_1,\omega_2,\mathbf{q}_2)= -\int_{0}^{\infty}du\int_{0}^{\infty}dv \langle\com{\com{\rho_{\mathbf{q}_{12}}(u+v)}{\rho_{-\mathbf{q}_1}(v)}}{\rho_{-\mathbf{q}_2}(0)}\rangle_0e^{i\omega_{12}^+ u}e^{i\omega_2^+ v},
\end{equation}
and we have introduced 
\begin{equation}
\mathbf{q}_{12}=\mathbf{q}_1+\mathbf{q}_2
\end{equation}
by analogy with Eq.~\eqref{eq:w12}. 
As in the general case Eq.~\eqref{eq:chi2physical}, the experimentally-measurable response function is not directly $\chi^{(2)}_{\rho\rho}(\omega_1,\mathbf{q}_1,\omega_2,\mathbf{q}_2)$, but is instead the symmetrized
\begin{align}
\bar{\chi}^{(2)}_{\rho\rho}(\omega_1,\mathbf{q}_1,\omega_2,\mathbf{q}_2) &= \frac{1}{2}\left(\chi^{(2)}_{\rho\rho}(\omega_1,\mathbf{q}_1,\omega_2,\mathbf{q}_2)+\chi^{(2)}_{\rho\rho}(\omega_2,\mathbf{q}_2,\omega_1,\mathbf{q}_1)\right).
\end{align}
Following the logic of Eq.~\eqref{eq:chi2specdensityrep}, we can introduce a spectral representation
\begin{equation}\label{eq:chi2rhospecdensityrep}
\chi^{(2)}_{\rho\rho}(\omega_1,\mathbf{q}_1,\omega_2,\mathbf{q}_2) = \frac{1}{\pi^2}\int  d\alpha d\beta \frac{\chi^{(2)''}_{\rho\rho}(\alpha,\mathbf{q}_1,\beta,\mathbf{q}_2)}{(\alpha-\omega_{12}^+)(\beta-\omega_2^+)},
\end{equation}
with the spectral density $\chi^{(2)''}_{\rho\rho}(\alpha,\mathbf{q}_1,\beta,\mathbf{q}_2)$ defined as the Fourier transform
\begin{equation}\label{eq:chi2rhospecdensity}
\chi^{(2)''}_{\rho\rho}(\alpha,\mathbf{q}_1,\beta,\mathbf{q}_2) = \frac{1}{4}\int  du dv \langle\com{\com{\rho_{\mathbf{q}_{12}}(u+v)}{\rho_{-\mathbf{q}_1}(v)}}{\rho_{-\mathbf{q}_2}(0)}\rangle_0e^{i(\alpha u + \beta v)}.
\end{equation}
\end{widetext}
Let us now consider second harmonic generation response $\omega_1=\omega_2=\omega$ to apply our sum rules Eqs.~\eqref{eq:shgasymptotic}--\eqref{eq:shgnu2}. 
To begin, we can define the second harmonic generation response function
\begin{align}
\bar{\chi}^{\mathrm{SHG}}_{\rho\rho}(\omega,\mathbf{q}_1,\mathbf{q}_2) &= \bar{\chi}^{(2)}_{\rho\rho}(\omega,\mathbf{q}_1,\omega,\mathbf{q}_2) \nonumber \\
&= \frac{1}{2}\left(\chi^{(2)}_{\rho\rho}(\omega,\mathbf{q}_1,\omega,\mathbf{q}_2)\right.\nonumber \\
&\left.+\chi^{(2)}_{\rho\rho}(\omega,\mathbf{q}_2,\omega,\mathbf{q}_1)\right). 
\end{align}
We can expand the spectral density representation Eq.~\eqref{eq:chi2rhospecdensityrep} to obtain an asymptotic series for $\bar{\chi}^{\mathrm{SHG}}_{\rho\rho}(\omega\rightarrow\infty,\mathbf{q}_1,\mathbf{q}_2)$ in the limit of large $\omega$. 
In analogy with Eq.~\eqref{eq:shgasymptotic} for the general response function, we find
\begin{equation}\label{eq:rhoshgasymptotic}
\bar{\chi}^{\mathrm{SHG}}_{\rho\rho}(\omega\rightarrow\infty,\mathbf{q}_1,\mathbf{q}_2)\sim\sum_{N=0}^\infty \frac{\nu_{\rho\rho,\mathbf{q}_1,\mathbf{q}_2}^{(N)}+\nu_{\rho\rho,\mathbf{q}_2,\mathbf{q}_1}^{(N)}}{4\omega^{N+2}},
\end{equation}
where the moments $\nu_{\rho\rho,\mathbf{q}_1,\mathbf{q}_2}^{(N)}$ are defined via
\begin{align}
\nu_{\rho\rho,\mathbf{q}_1,\mathbf{q}_2}^{(N)} = \frac{1}{\pi^2}\sum_{m=0}^{N}\int d\alpha d\beta &\left(\frac{\alpha}{2}\right)^{N-m}\beta^m \nonumber \\
&\times\chi^{(2)''}_{\rho\rho}(\alpha,\mathbf{q}_1,\beta,\mathbf{q}_2).
\end{align}
 For the zeroth moment $\nu^{(0)}_{\rho\rho,\mathbf{q}_1,\mathbf{q}_2}$ we find
\begin{align}\label{eq:shgrhonu0}
\nu^{(0)}_{\rho\rho,\mathbf{q}_1,\mathbf{q}_2} & =  \frac{1}{\pi^2}\int d\alpha d\beta\chi^{(2)''}_{\rho,\rho}(\alpha,\mathbf{q}_1,\beta,\mathbf{q}_2) \nonumber\\
&=\langle\com{\com{\rho_{\mathbf{q}_{12}}}{\rho_{-\mathbf{q}_1}}}{\rho_{-\mathbf{q}_2}}\rangle_0 \nonumber\\
&=0,
\end{align}
since all Fourier components of the density operator are mutually commuting. 
For the first moment $\nu^{(1)}_{\rho\rho, \mathbf{q}_1,\mathbf{q}_2}$ we find from Eq.~\eqref{eq:shgnu1} that
\begin{align}\label{eq:shgnu1intermsofrhos}
\nu^{(1)}_{\rho\rho, \mathbf{q}_1,\mathbf{q}_2} & =  \frac{1}{\pi^2}\int d\alpha d\beta\left(\frac{\alpha}{2}+\beta\right)\chi^{(2)''}_{\rho,\rho}(\alpha,\mathbf{q}_1,\beta,\mathbf{q}_2) \nonumber\\
&=\frac{1}{2}\langle\com{\com{i\partial_t \rho_{\mathbf{q}_{12}}}{ \rho_{-\mathbf{q}_1} }}{ \rho_{-\mathbf{q}_2} }\rangle_0,
\end{align}
where we have again made use of the fact that density operators have vanishing equal-time commutators. 

We can additionally use charge conservation to re-express Eq.~\eqref{eq:shgnu1intermsofrhos} in terms of a variation of the unperturbed Hamiltonian $H_0$ with respect to the electromagnetic vector potential. 
As we review in Appendix~\ref{sec:rhorho1}, conservation of charge along with the canonical commutation relations imply that for any operator $\mathcal{O}$,
\begin{equation}\label{eq:gaugevariationmain}
\left[\mathcal{O},\rho_\mathbf{q}\right] =\left.q^\mu\frac{\delta\mathcal{O}_A}{\delta A^\mu_\mathbf{-q}}\right|_{\mathbf{A}\rightarrow 0}, 
\end{equation}
where $\mathcal{O}_A$ is the operator $\mathcal{O}$ minimally coupled to the external vector potential $\mathbf{A}_\mathbf{q}$.
We find
\begin{align}\label{eq:shgrhonu1}
\nu^{(1)}_{\rho\rho, \mathbf{q}_1,\mathbf{q}_2}  &= \frac{1}{2}(q_{12}^\mu)(q_1^\nu)(q_2^\lambda)\nonumber \\
&\times\left.\left\langle\frac{\delta^3 H}{\delta A^\mu_{-\mathbf{q}_{12}}\delta A^\nu_{\mathbf{q}_1}\delta A^\lambda_{\mathbf{q}_2}}\right\rangle_0\right|_{\mathbf{A}\rightarrow 0},
\end{align}
The third variation of the Hamiltonian with respect to the vector potential gives the generalized diamagnetic current, describing corrections to the current operator that are quadratic in the vector potential. 
These are sometimes known as ``three photon vertices''~\cite{parker2019diagrammatic}. 
As shown in Ref.~\cite{mckay2023spatially}, the right hand side of Eq.~\eqref{eq:shgrhonu1} can be evaluated as a nested commutator between the position operator and the ordinary (paramagnetic) current operator.

Combining Eqs.~\eqref{eq:shgrhonu0} and \eqref{eq:shgrhonu1}, we see that the second harmonic generation response function satisfies
\begin{align}\label{eq:shg3photon}
\lim_{\omega\rightarrow\infty }\omega^3&\bar{\chi}^{\mathrm{SHG}}_{\rho\rho}(\omega,\mathbf{q}_1,\mathbf{q}_2) =  \\
&\frac{1}{4}(q_{12}^\mu)(q_1^\nu)(q_2^\lambda)\left.\left\langle\frac{\delta^3 H}{\delta A^\mu_{-\mathbf{q}_{12}}\delta A^\nu_{\mathbf{q}_1}\delta A^\lambda_{\mathbf{q}_2}}\right\rangle_0\right|_{\mathbf{A}\rightarrow 0}\nonumber.
\end{align}
Eqs.~\eqref{eq:shgrhonu1} and \eqref{eq:shg3photon} give the first generalization of the linear $f$-sum rule [see Eq.~\eqref{eq:fsumasymptotic}] to the second-order density response, and are a main result of this note.

However, for nonrelativistic systems, the Hamiltonian is at most quadratic in momentum and interactions are momentum-independent. 
We can write a general nonrelativistic Hamiltonian as
\begin{equation}\label{eq:hnr}
H_{\mathrm{n.r.}}=\sum_i \frac{|\mathbf{p}_i|^2}{2m} - \vec{\lambda(\mathbf{x}_i)}\cdot(\mathbf{p}\times\vec\sigma) + V(\mathbf{x}_i) + \frac{1}{2}\sum_{i\neq j} U(\mathbf{x}_i-\mathbf{x}_j),
\end{equation}
where $i,j$ index particles in the system, $m$ is the electron mass, $V$ is an external potential, $\vec{\lambda}$ is the spin-orbit potential, $\vec{\sigma}$ is a vector of Pauli matrices acting on the electron spins, and $U$ is a pair interaction. 
Minimally coupling $H_{\mathbf{n.r.}}$ to a vector potential via $\mathbf{p}_i\rightarrow \mathbf{p}_i - \mathbf{A}(\mathbf{x}_i)$ produces terms of at-most quadratic order in the vector potential $\mathbf{A}$. 
As such, we have
\begin{equation}
\left.\frac{\delta^3 H_{\mathrm{n.r.}}}{\delta A^\mu_{-\mathbf{q}_{12}}\delta A^\nu_{\mathbf{q}_1}\delta A^\lambda_{\mathbf{q}_2}}\right|_{\mathbf{A}\rightarrow 0} = 0.
\end{equation}
Thus, for systems governed by nonrelativistic Hamiltonians of the form of Eq.~\eqref{eq:hnr}, $\nu^{(1)}_{\rho\rho, \mathbf{q}_1,\mathbf{q}_2}$ vanishes identically, and we need to go to the second moment to obtain a nontrivial sum rule and nonzero asymptotic expansion. 

Turning to $\nu^{(2)}_{\rho\rho,\mathbf{q}_1,\mathbf{q}_2}$ we have
\begin{align}\label{eq:shgrhonu2part1}
&\nu^{(2)}_{\rho\rho,\mathbf{q}_1,\mathbf{q}_2} = \frac{1}{\pi^2}\int d\alpha d\beta\left(\frac{\alpha^2}{4}+\beta^2+\frac{\alpha\beta}{2}\right)\chi^{(2)''}_{\rho\rho}(\alpha,\mathbf{q}_1,\beta,\mathbf{q}_2) \nonumber\\
&= -\frac{2\langle\com{\partial_t\com{\partial_t \rho_{\mathbf{q}_{12}}}{\rho_{-\mathbf{q}_1}}}{\rho_{-\mathbf{q}_2}}\rangle_0 +\langle\com{\com{\partial^2_t \rho_{\mathbf{q}_{12}}}{\rho_{-\mathbf{q}_1}}}{\rho_{-\mathbf{q}_2}}\rangle_0}{4}\nonumber \\
&= \frac{-3\langle\com{\partial_t\com{\partial_t \rho_{\mathbf{q}_{12}}}{\rho_{-\mathbf{q}_1}}}{\rho_{-\mathbf{q}_2}}\rangle_0 +\langle\com{\com{\partial_t \rho_{\mathbf{q}_{12}}}{\partial_t\rho_{-\mathbf{q}_1}}}{\rho_{-\mathbf{q}_2}}\rangle_0}{4}.
\end{align}
Eq.~\eqref{eq:shgrhonu2part1} is general, and requires no assumptions on the form of the Hamiltonian. 
However, for nonrelativistic systems we can simplify it further by making use of Eqs.~\eqref{eq:gaugevariationmain} and Eq.~\eqref{eq:hnr} to find that for nonrelativistic systems the density algebra satisfies
\begin{equation}
\com{\partial_t \rho_\mathbf{q} } {\rho_{\mathbf{q'}}}\rightarrow_{\mathrm{n.r.}} i\frac{\mathbf{q}\cdot\mathbf{q'}}{m}\rho_{\mathbf{q+q'}}. 
\end{equation}
By using the Jacobi identity to rewrite the second term in Eq.~\eqref{eq:shgrhonu2part1} solely in terms of commutators of the density with time derivatives of the density, we find that in the nonrelativistic limit
\begin{align}\label{eq:nu2nr}
\nu^{(2)}_{\rho\rho,\mathbf{q}_1,\mathbf{q}_2}&\rightarrow_{\mathrm{n.r.}} \frac{3 i\mathbf{q}_{12}\cdot \mathbf{q}_1 }{4m} \langle \com {\partial_t \rho_{\mathbf{q}_2} }{\rho_{-\mathbf{q}_2}}\rangle_0\nonumber \\
& -\frac{1}{4}\langle \com{\com{ \partial_t\rho_\mathbf{-q_1} }{ \rho_{\mathbf{-q_2}} }}{\partial_t\rho_{\mathbf{q}_{12}} }\rangle_0 \nonumber \\
&+\frac{1}{4}\langle \com{\com{ \partial_t\rho_\mathbf{q_{12}} }{ \rho_{\mathbf{-q_2}} }}{\partial_t\rho_\mathbf{-q_1} }\rangle_0 \nonumber\\
&=\frac{\bar{n}\mathbf{q}_{12}\cdot\left[3 \mathbf{q}_1 |\mathbf{q}_2|^2 +\mathbf{q}_{12}(\mathbf{q}_1\cdot \mathbf{q}_2) + \mathbf{q}_2 |\mathbf{q}_1|^2\right] }{4m^2},
\end{align}
where $\bar{n}=\langle\rho_\mathbf{0}\rangle_0$ is the average charge density in equilibrium.

Substituting Eq.~\eqref{eq:nu2nr} into our asymptotic expansion Eq.~\eqref{eq:rhoshgasymptotic}, we see that, in the nonrelativistic limit the second harmonic generation response at large frequency is asymptotically given by
\begin{align}
%\bar{\chi}^{\mathrm{SHG}}_{\rho\rho}(\omega\rightarrow\infty,\mathbf{q}_1,\mathbf{q}_2)\sim_{\mathrm{n.r.}}\frac{\bar{n}}{8m^2\omega^4}\left(4|\mathbf{q}_1|^2|\mathbf{q}_2|^2+2(\mathbf{q}_1\cdot\mathbf{q}_2)^2+3\mathbf{q}_1\cdot\mathbf{q}_2(|\mathbf{q}_1|^2+|\mathbf{q}_2|^2\right) +\mathcal{O}(\omega^{-5})
\bar{\chi}^{\mathrm{SHG}}_{\rho\rho}(\omega\rightarrow\infty,\mathbf{q}_1,\mathbf{q}_2)&\sim_{\mathrm{n.r.}}\frac{\bar{n}}{8m^2\omega^4}\Big(4|\mathbf{q}_1|^2|\mathbf{q}_2|^2\Big.\nonumber \\
&\left.-4(\mathbf{q}_1\cdot\mathbf{q}_2)^2\right.\nonumber \\&\left.+3\mathbf{q}_1\cdot\mathbf{q}_2|\mathbf{q}_{12}|^2\right) +\mathcal{O}(\omega^{-5}).
\end{align}
Finally, note that for a harmonic perturbation, $f_\mathbf{q}$ has only two nonvanishing Fourier components at $\mathbf{q}=\pm \mathbf{q_0}$. 
In this case, the relevant moments are
\begin{equation}
\nu^{(2)}_{\rho\rho}(\mathbf{q}_0,-\mathbf{q}_0) \rightarrow_{\mathrm{n.r.}} 0
\end{equation}
and
\begin{equation}\label{eq:shgfsumnr1}
\nu^{(2)}_{\rho\rho}(\mathbf{q}_0,\mathbf{q}_0) \rightarrow_{\mathrm{n.r.}} \frac{3|\mathbf{q_0}|^4\bar{n}}{m^2}
\end{equation}
and hence
\begin{align}
\bar{\chi}^{\mathrm{SHG}}_{\rho\rho}(\omega\rightarrow\infty,\mathbf{q}_0,-\mathbf{q}_0)&\sim_{\mathrm{n.r.}} 0, \\
\bar{\chi}^{\mathrm{SHG}}_{\rho\rho}(\omega\rightarrow\infty,\mathbf{q}_0,\mathbf{q}_0)&\sim_{\mathrm{n.r.}} \frac{3\bar{n}|\mathbf{q}_0|^4}{2m^2\omega^4},\label{eq:shgfsumnr2}
\end{align}
Note that $\bar{\chi}^{\mathrm{SHG}}_{\rho\rho}(\omega,\mathbf{q}_0,-\mathbf{q}_0)$ determines the second order response 
\begin{equation}
\delta^{2}\langle\rho_\mathbf{0}\rangle(2\omega) = \bar{\chi}^{\mathrm{SHG}}_{\rho\rho}(\omega,\mathbf{q}_0,-\mathbf{q}_0)f_{\mathbf{q}_0}(\omega)f_{-\mathbf{q}_0}(\omega).
\end{equation}
In fact, $\bar{\chi}^{\mathrm{SHG}}_{\rho\rho}(\omega,\mathbf{q}_0,-\mathbf{q}_0)$ is identically zero for systems with particle number conservation, since in that case the $\mathbf{q}=0$ Fourier component of the density operator is time independent, so the commutator in Eq.~\eqref{eq:chi2specdensity} vanishes.

Similarly, $\bar{\chi}^{\mathrm{SHG}}_{\rho\rho}(\omega,\mathbf{q}_0,\mathbf{q}_0)$ determines the second-order response
\begin{equation}
\delta^{2}\langle\rho_{2\mathbf{q}_0}\rangle(2\omega) = \bar{\chi}^{\mathrm{SHG}}_{\rho\rho}(\omega,\mathbf{q}_0,\mathbf{q}_0)f_{\mathbf{q}_0}(\omega)f_{\mathbf{q}_0}(\omega).
\end{equation}
Thus, the nonrelativistic sum rules Eqs.~\eqref{eq:shgfsumnr1} and \eqref{eq:shgfsumnr2} constrain the large-frequency asymptotic response of the density Fourier component $\langle\rho_{2\mathbf{q}_0}\rangle(2\omega)$ in the limit of large frequency. 
These sum rules place important constraints on, and should aid the interpretation of, the newest generation of nonlinear x-ray optics experiments, which exploit second-order optical effects at large momenta and have potential new applications in surface science, low-dimensional transport, and multidimensional spectroscopies involving core levels \cite{Chergui2023,Rohlsberger2020}. 

\section{Example: Density Response for Free fermions in a crystal}\label{sec:freefermions}

We will now see how our density response sum rules constrain the properties of noninteracting electrons in solids. 
We consider an unperturbed Hamiltonian $H_0$ with discrete translation symmetry consisting solely of one-particle terms. 
We can label the single particle eigenstates $\ket{\psi_{n\mathbf{k}}}$ of $H_0$ by their crystal momentum $\mathbf{k}$ in the first Brillouin zone, such that
\begin{equation}\label{eq:blocheigs}
H_0\ket{\psi_{n\mathbf{k}}} = \epsilon_{n\mathbf{k}}\ket{\psi_{n\mathbf{k}}}.
\end{equation}
It will be computationally convenient to recast Eq.~\eqref{eq:blocheigs} in terms of the cell-periodic functions
\begin{equation}
\ket{u_{n\mathbf{k}}} = e^{-i\mathbf{k}\cdot\mathbf{x}}\ket{\psi_{n\mathbf{k}}},
\end{equation}
where $\mathbf{x}$ is the single-particle position operator. 
Introducing the Bloch Hamiltonian
\begin{equation}
H_\mathbf{k} = e^{-i\mathbf{k}\cdot\mathbf{x}}H_0e^{i\mathbf{k}\cdot\mathbf{x}},
\end{equation}
we have
\begin{equation}
H_\mathbf{k}\ket{u_{n\mathbf{k}}} = \epsilon_{n\mathbf{k}}\ket{u_{n\mathbf{k}}}.
\end{equation}
We will normalize the $\ket{u_{n\mathbf{k}}}$ over a single unit cell. 
Introducing creation and annihilation operators $c^\dag_{n\mathbf{k}}, c_{n\mathbf{k}}$ corresponding to the states $\ket{\psi_{n\mathbf{k}}}$, we can write the Hamiltonian $H_0$ in second quantization as
\begin{equation}\label{eq:h02nd}
H_0 = \sum_{n\mathbf{k}}\epsilon_{n\mathbf{k}}c^\dag_{n\mathbf{k}}c_{n\mathbf{k}}.
\end{equation}
Similarly, the density operator in second quantization is given by
\begin{equation}\label{rho2nd}
\rho_\mathbf{q} = \sum_{nm\mathbf{k}}\braket{u_{n\mathbf{k}}}{u_{m\mathbf{k+q}}}c^\dag_{n\mathbf{k}}c_{m\mathbf{k+q}},
\end{equation}
where the overlap between cell-periodic functions is defined as an integral over one unit cell. 

For a free-fermion ground state, we can directly evaluate the spectral density Eq.~\eqref{eq:chi2specdensity} for the second-order density response function as a sum of density operator matrix elements times Fermi-Dirac distribution functions, using Eq.~\eqref{eq:specdensitydelta}. 
The expression is exact, but unwieldy. 
On the other hand, we can use Eqs.~\eqref{eq:h02nd} and \eqref{rho2nd} along with canonical anticommutation relations to derive a simple exact expression for the nonlinear sum rule Eq.~\eqref{eq:shgnu1intermsofrhos}. 
To do so, we first use the equations of motion to rewrite Eq.~\eqref{eq:shgnu1intermsofrhos} as
\begin{equation}
\nu^{(1)}_{\rho\rho,\mathbf{q}_1,\mathbf{q}_2} = -\frac{1}{2}\expval{\com{
\com{
\com{H} {\rho_{\mathbf{q}_{12}}}
} 
{\rho_{-\mathbf{q}_1}}
}
{\rho_{-\mathbf{q}_2}}
}_0.\label{eq:nu1comm}
\end{equation}
To evaluate the nested commutators, we can make use of the following relation: Given any single-body operator
\begin{equation}
B_\mathbf{q}=\sum_{\mathbf{k}nm} \mel{u_{n\mathbf{k}}}{B_\mathbf{k,q}}{u_{m\mathbf{k+q}}}c^\dag_{n\mathbf{k}}c_{n\mathbf{k+q}},
\end{equation}
we can use the canonical anticommutation relations to find
\begin{equation}\label{eq:rhocomm}
[B_\mathbf{q},\rho_\mathbf{q'}] = \sum_{\mathbf{k}nm}\mel{u_{n\mathbf{k}}}{B_\mathbf{k,q}-B_\mathbf{k+q',q}}{u_{m\mathbf{k+q+q'}}}.
\end{equation}
Making extensive use of Eq.~\eqref{eq:rhocomm} to evaluate the nested commutators in Eq.~\eqref{eq:nu1comm}, we find that
\begin{widetext}
\begin{align}
\com{
\com{
\com{H} {\rho_{\mathbf{q}_{12}}}
} 
{\rho_{-\mathbf{q}_1}}
}
{\rho_{-\mathbf{q}_2}} & = \sum_{nm\mathbf{k}}c^\dag_{n\mathbf{k}}c_{m\mathbf{k}}\mel**{u_{n\mathbf{k}}}{H_{\mathbf{k}-\mathbf{q}_1-\mathbf{q}_2} + H_{\mathbf{k}+\mathbf{q}_1}+H_{\mathbf{k}+\mathbf{q}_2}-H_{\mathbf{k}+\mathbf{q}_1+\mathbf{q}_2}-H_{\mathbf{k}-\mathbf{q}_1}-H_{\mathbf{k}-\mathbf{q}_1}}{u_{m\mathbf{k}}}.
\end{align}

We can evaluate this average in a thermal ground state with fixed temperature $T$ and chemical potential $\mu$. 
Let $f_{n\mathbf{k}}$ be the Fermi-Dirac distribution function evaluated at energy $\epsilon_{n\mathbf{k}}$. 
Then we have
\begin{align}\label{eq:freefermionnu1}
\nu^{(1)}_{\rho\rho,\mathbf{q}_1,\mathbf{q}_2} &= \frac{1}{2}\sum_{n\mathbf{k}}f_{n\mathbf{k}}\mel**{u_{n\mathbf{k}}}{H_{\mathbf{k}+\mathbf{q}_1+\mathbf{q}_2} + H_{\mathbf{k}-\mathbf{q}_1}+H_{\mathbf{k}-\mathbf{q}_2}-H_{\mathbf{k}-\mathbf{q}_1-\mathbf{q}_2}-H_{\mathbf{k}+\mathbf{q}_1}-H_{\mathbf{k}+\mathbf{q}_1}}{u_{n\mathbf{k}}}.
\end{align}
\end{widetext}
Eq.~\eqref{eq:freefermionnu1} expresses the first moment of the second order spectral density entirely in terms of matrix elements of the Bloch Hamiltonian. 
For low-energy Wannier-based models of condensed matter systems, Eq.~\eqref{eq:freefermionnu1} combined with Eq.~\eqref{eq:shgrhonu1} shows that there is an effective second-order diamagnetic current proportional to $\nu^{(1)}_{\rho\rho,\mathbf{q}_1,\mathbf{q}_2}$. 
Furthermore, we can use the Karplus-Schwinger relation of Refs.~\cite{karplus1948note,mckay2023spatially} to rewrite Eq.~\eqref{eq:freefermionnu1} in terms of derivatives of the Bloch Hamiltonian as
\begin{align}\label{eq:karplusschwingernu1}
\nu^{(1)}_{\rho\rho,\mathbf{q}_1,\mathbf{q}_2} =& q_{12}^\mu q_1^\nu q_2^\lambda\frac{1}{2}\Bigg[\sum_{n\mathbf{k}}\int_0^1d\lambda\int_0^1d\lambda_1\int_0^1d\lambda_2 f_{n\mathbf{k}}\Bigg.\nonumber \\
&\Bigg.\times \mel**{u_{n\mathbf{k}}}{\frac{\partial^3 H_{\mathbf{k}-\lambda\mathbf{q}_{12}+\lambda_1\mathbf{q}_1+\lambda_2\mathbf{q}_2}}{\partial k^\mu\partial k^\nu\partial k^\lambda}}{u_{n\mathbf{k}}}\Bigg],
\end{align}
which shows that the longitudinal component of the diamagnetic current coincides with the longitudinal component of the term in brackets in Eq.~\eqref{eq:karplusschwingernu1}. 
This is consistent with the results of Ref.~\cite{mckay2023spatially}.

\section{Second-order Conductivity sum rule}\label{sec:cond}

From charge conservation, it follows that the second order density response function $\chi^{(2)}_{\rho\rho}(\mathbf{q}_1,\omega_1,\mathbf{q}_2,\omega_2)$ determines the longitudinal components of the second order conductivity tensor, and hence our sum rules in Eqs.~\eqref{eq:chi2asymptotic} and \eqref{eq:shgnu0}--\eqref{eq:shgnu2} for the density response imply longitudinal second order conductivity sum rules. 
To see this concretely, let us return to the general formalism of Sec.~\ref{sec:2ndorder}. 
We consider again a scalar potential perturbation
\begin{equation}\label{eq:phipert}
H_1=\sum_{\mathbf{q}}\phi_\mathbf{q}(t)\rho_{-\mathbf{q}},
\end{equation}
and look at the response of the current operator $\mathbf{j}_{\mathbf{q}}$. 
At second order in the scalar potential, we have
\begin{widetext}
\begin{equation}
\delta^2\langle j^\mu_\mathbf{q}\rangle(\omega) = \sum_{\mathbf{q}_1,\mathbf{q}_2}\int d\omega_1 d\omega_2\chi^{(2),\mu}_{j\rho}(\omega_1,\mathbf{q}_1,\omega_2,\mathbf{q}_2)\delta(\mathbf{q}-\mathbf{q}_{12})\delta(\omega-\omega_{12})\phi_{\mathbf{q}_1}(\omega_1)\phi_{\mathbf{q}_2}(\omega_2),
\end{equation}
where the nonlinear Kubo formula gives
\begin{equation}\label{eq:chijrho}
\chi^{(2),\mu}_{j\rho}(\omega_1,\mathbf{q}_1,\omega_2,\mathbf{q}_2)= -\int_{0}^{\infty}du\int_{0}^{\infty}dv \langle\com{\com{j^\mu_{\mathbf{q}_{12}}(u+v)}{\rho_{-\mathbf{q}_1}(v)}}{\rho_{-\mathbf{q}_2}(0)}\rangle_0e^{i\omega_{12}^+ u}e^{i\omega_2^+ v},
\end{equation}
However, from gauge invariance, we know that the current cannot respond to the scalar potential directly, but only to the electric field. 
In the gauge of Eq.~\eqref{eq:phipert} where the vector potential is zero, the electric field is given by
\begin{equation}
\mathbf{E}_\mathbf{q}(t) = -i\mathbf{q}\phi_\mathbf{q}(t).
\end{equation}
Defining the nonlinear conductivity $\sigma^{\mu\nu\lambda}(\omega_1,\mathbf{q}_1,\omega_2,\mathbf{q}_2)$ via the constitutive relation
\begin{equation}
\delta^2\langle j^\mu_\mathbf{q}\rangle(\omega)=\sum_{\mathbf{q,q'}}\int d\omega_1 d\omega_2\sigma^{\mu\nu\lambda}(\omega_1,\mathbf{q}_1,\omega_2,\mathbf{q}_2)\delta(\mathbf{q}-\mathbf{q}_{12})\delta(\omega-\omega_{12})E^\nu_{\mathbf{q}_1}(\omega_1)E^\lambda_{\mathbf{q}_2}(\omega_2),
\end{equation}
\end{widetext}
We see that we can choose~\footnote{There is an ambiguity here due to the fact that only the response functions symmetrized under $\mathbf{q}_1\leftrightarrow\mathbf{q}_2$ are determined from the constitutive relations. 
Nevertheless, we are free to make any choice we wish for the antisymmetric terms provided we symmetrize before obtaining physical results.}
\begin{equation}\label{eq:chisigmarel}
\chi^{(2),\mu}_{j\rho}(\omega_1,\mathbf{q}_1,\omega_2,\mathbf{q}_2)=-q_1^\nu q_2^\lambda\sigma^{\mu\nu\lambda}(\omega_1,\mathbf{q}_1,\omega_2,\mathbf{q}_2).
\end{equation}
To progress further, we can contract $\chi^{(2),\mu}_{j\rho}$ with $\mathbf{q}_{12}$ and use the continuity equation
\begin{equation}\label{eq:cont}
i\partial_t\rho_{\mathbf{q}_{12}} = \mathbf{q}_{12}\cdot\mathbf{j}_{\mathbf{q}_{12}}
\end{equation}
to relate $\sigma^{\mu\nu\lambda}(\omega_1,\mathbf{q}_1,\omega_2,\mathbf{q}_2)$ with the density response function $\chi^{(2)}_{\rho\rho}(\omega_1,\mathbf{q}_1,\omega_2,\mathbf{q}_2)$ from Eq.~\eqref{eq:chirho2kubo}. 
Inserting Eq.~\eqref{eq:cont} into Eq.~\eqref{eq:chijrho} and integrating by parts we find
\begin{equation}\label{eq:condchi}
-q_{12}^\mu q_1^\nu q_2^\lambda\sigma^{\mu\nu\lambda}(\omega_1,\mathbf{q}_1,\omega_2,\mathbf{q}_2) = \omega_{12}^+\chi^{(2)}_{\rho\rho}(\omega_1,\mathbf{q}_1,\omega_2,\mathbf{q}_2),
\end{equation}
Eq.~\eqref{eq:condchi} relates the longitudinal second-order conductivity to the second-order density response function. 
Going further, we can make use of the spectral representation Eq.~\eqref{eq:chi2specdensityrep} to express the longitudinal second-order conductivity as 
\begin{align}\label{eq:sigmaspec}
-q_{12}^\mu q_1^\nu q_2^\lambda&\sigma^{\mu\nu\lambda}(\omega_1,\mathbf{q}_1,\omega_2,\mathbf{q}_2) \nonumber \\
&= \frac{1}{\pi^2}\int  d\alpha d\beta \frac{\alpha\chi^{(2)''}_{\rho\rho}(\alpha,\mathbf{q}_1,\beta,\mathbf{q}_2)}{(\alpha-\omega_{12}^+)(\beta-\omega_2^+)} \\
&=-\frac{1}{\pi^2}\int  d\alpha d\beta \frac{\sigma_L''(\alpha,\mathbf{q}_1,\beta,\mathbf{q}_2)}{(\alpha-\omega_{12}^+)(\beta-\omega_2^+)}.
\end{align}
We see then that the spectral density $\sigma_L''(\alpha,\mathbf{q}_1,\beta,\mathbf{q}_2)$ for the longitudinal second-order conductivity is given by $-\alpha$ times the spectral density $\chi^{(2)''}_{\rho\rho}(\alpha,\mathbf{q}_1,\beta,\mathbf{q}_2)$ for the second-order density response function. 
We can now follow the logic of Sec.~\ref{sec:sumrule} to asymptotically expand Eq.~\eqref{eq:sigmaspec} at large frequency to deduce sum rules for the longitudinal conductivity. 
Focusing on the second harmonic current $\omega_1=\omega_2=\omega$ at large $\omega$ and explicitly symmetrizing under $\mathbf{q}_1\leftrightarrow \mathbf{q}_2$ to obtain the physical response [we denote the symmetrization with an overbar, extending the notation of Eq.~\eqref{eq:chi2physical}]
\begin{equation}
q_{12}^\mu q_1^\nu q_2^\lambda\bar{\sigma}^{\mu\nu\lambda}(\omega,\mathbf{q}_1,\omega,\mathbf{q}_2) \sim \sum_{N=0}^\infty\frac{\mu^{(N)}_{\mathbf{q}_1,\mathbf{q}_2}+\mu^{(N)}_{\mathbf{q}_2,\mathbf{q}_1}}{4\omega^{N+2}},
\end{equation}
where the moments $\mu^{(N)}_{\mathbf{q}_1,\mathbf{q}_2}$ are defined as
\begin{align}
\mu^{(N)}_{\mathbf{q}_1,\mathbf{q}_2}&=\frac{1}{\pi^2}\sum_{m=0}^{N}\int d\alpha d\beta \left(\frac{\alpha}{2}\right)^{N-m}\beta^m\sigma_L''(\alpha,\mathbf{q}_1,\beta,\mathbf{q}_2) \nonumber \\
&=\frac{1}{\pi^2}\sum_{m=0}^{N}\int d\alpha d\beta \left(\frac{\alpha}{2}\right)^{N-m}\beta^m%\nonumber \\ &\times
\alpha\chi^{(2)''}_{\rho\rho}(\alpha,\mathbf{q}_1,\beta,\mathbf{q}_2).
\end{align}
In particular, we find that
\begin{align}
\mu^{(0)}_{\mathbf{q}_1,\mathbf{q}_2}&=2\nu^{(1)}_{\rho\rho, \mathbf{q}_1,\mathbf{q}_2}\label{eq:mu0}, \\
\mu^{(1)}_{\mathbf{q}_1,\mathbf{q}_2}&=2\nu^{(2)}_{\rho\rho,\mathbf{q}_1,\mathbf{q}_2}\label{eq:mu1},
\end{align}
where $\nu^{(1)}_{\rho\rho, \mathbf{q}_1,\mathbf{q}_2}$ and $\nu^{(2)}_{\rho\rho,\mathbf{q}_1,\mathbf{q}_2}$ are the moments defined in Eqs.~\eqref{eq:shgnu1} and \eqref{eq:shgnu2}, respectively. 
Eqs.~\eqref{eq:mu0} and \eqref{eq:mu1} are sum rules for the longitudinal second order conductivity. 
In particular, Eq.~\eqref{eq:shgnu1} shows that $\mu^{(0)}_{\mathbf{q}_1\mathbf{q}_2}$ is proportional to the average third derivative of the Hamiltonian with respect to the vector potential. 
This is equivalent to the instantaneous sum rule for the second-order conductivity derived in Ref.~\cite{watanabe2020generalized}. 
Our formalism extends this result beyond leading asymptotic order in $\omega$, which is particularly important for nonrelativistic systems in which $\nu^{(1)}_{\rho\rho, \mathbf{q}_1,\mathbf{q}_2}$ is identically zero. 
Finally, applied to free-fermion systems, we see from Eqs.~\eqref{eq:freefermionnu1} and \eqref{eq:karplusschwingernu1} that the conductivity sum rule Eq.~\eqref{eq:mu0} can be written entirely in terms of matrix elements of the Bloch Hamiltonian (or of its third derivative).

\section{Rectification Sum Rule}\label{sec:rect}
Up to now, we have focused primarily on second-harmonic generation response $\omega_1=\omega_2$. 
However, for response to real harmonic perturbations at second order, there is also a rectification response with $\omega_1=-\omega_2$. 
The rectification response determines the zero frequency (time-independent) response of the average $\langle A \rangle(\omega=0)$ to the perturbation $f(t)$ at second order; setting $\omega=0$ in Eq.~\eqref{eq:chi2deffourier}, we can defined the rectification response $\chi_{AB}^\mathrm{R}(\omega)$ via
\begin{equation}
\langle A \rangle(\omega=0) = \int d\omega \chi_{AB}^\mathrm{R}(\omega)|f(\omega)|^2,
\end{equation}
where we have used the fact that for real perturbations $f(-\omega)=f(\omega)^*$. 
Comparing with Eq.~\eqref{eq:chi2deffourier} we have that
\begin{equation}
\chi_{AB}^\mathrm{R}(\omega) = \frac{1}{2}\left(
\chi_{AB}^{(2)}(\omega,-\omega) + \chi_{AB}^{(2)}(-\omega,\omega)
\right).
\end{equation}
The spectral representation for the rectification response takes a particularly simple form. 
Setting $\omega_1=-\omega_2=\omega$ in the spectral representation Eq.~\eqref{eq:chi2specdensityrep} and symmetrizing, we find
\begin{equation}\label{eq:rectspectralrep}
\chi_{AB}^\mathrm{R}(\omega)= \frac{1}{\pi^2} \int d\alpha d\beta \frac{\beta\chi_{AB}^{(2)''}(\alpha,\beta)}{(\alpha-2i\epsilon)(\beta^2-(\omega^+)^2)},
\end{equation}
which defines the rectification response in terms of an integral of the spectral density.

Analogous to our treatment second-harmonic generation response in Sec.~\ref{sec:sumrule}, we can use Eq.~\eqref{eq:rectspectralrep} to derive an asymptotic expansion for the rectification response in the limit of large frequency $|\omega|\rightarrow\infty$. 
We find asymptotically that
\begin{equation}\label{eq:rectasymp}
\chi_{AB}^\mathrm{R}(\omega\rightarrow\infty)\sim\sum_{N=0}^{\infty}\frac{\zeta^{(N)}_{AB}}{2\omega^{2N+2}},
\end{equation}
where we have introduced the moments
\begin{equation}\label{eq:zetadef}
 \zeta^{(N)}_{AB} = -\frac{1}{\pi^2}\int d\alpha d\beta \frac{\beta^{2N+1}\chi_{AB}^{(2)''}(\alpha,\beta)}{\alpha-2i\epsilon}.
\end{equation}
Note that, unlike our sum rule for second harmonic generation, the moments $\zeta^{(N)}_{AB}$ involve the negative first moment in $\alpha$ of the spectral density. 
This is reminiscent of the sum rule for polarization fluctuations introduced in Refs.~\cite{souza2000polarization,onishi2024quantum,verma2024instantaneous}. 
In general, the presence of a negative power of $\alpha$ hinders our ability to simplify Eq.~\eqref{eq:zetadef} using our definition Eq.~\eqref{eq:chi2specdensity} of the spectral density. 
However, in the special case that $A$ can be written as a time derivative
\begin{equation}\label{eq:Aasderiv}
A=i[H_0,C]\equiv \partial_t C
\end{equation}
of some operator $C$ in the Heisenberg picture, we can combine Eqs.~\eqref{eq:zetadef}, \eqref{eq:Aasderiv} and Eq.~\eqref{eq:chi2specdensity} and integrate by parts to obtain
\begin{equation}\label{eq:zetasumrule}
    \zeta^{N}_{AB} = (-1)^{N}\expval{\com{\com{C}{B}}{\partial_t^{2N+1}B}}_0,
\end{equation}
where the time derivatives of $B$ are evaluated in the Heisenberg picture as commutators with $H_0$. 
Note that from Eq.~\eqref{eq:Aasderiv}, only the off-diagonal components of $C$ (in the eigenbasis of $H_0$) determine $A$; to avoid ambiguities with integration by parts, we take $C$ to be off-diagonal in the unperturbed energy eigenbasis. 
This is enforced by the $i\epsilon$ in the denominator of Eq.~\eqref{eq:zetadef}, which can be seen by inserting Eq.~\eqref{eq:specdensitydelta} into Eq.~\eqref{eq:zetadef}.

Eqs.~\eqref{eq:rectasymp}, \eqref{eq:zetadef}, and \eqref{eq:zetasumrule} give a family of generalized sum rules for the second order rectification response. 
While not directly applicable to density response (since $\rho_\mathbf{q}$ cannot in general be written as the time derivative of an operator), our sum rule can be applied to the spatially uniform ($\mathbf{q}\rightarrow 0$) conductivity, using the fact that the uniform current can be written as the time derivative of the (off-diagonal matrix elements of the) position operator. 
Thus Eqs.~\eqref{eq:rectasymp}, \eqref{eq:zetadef}, and \eqref{eq:zetasumrule} extends the rectification sum rule of Refs.~\cite{matsyshyn2019nonlinear,matsyshyn2021berry}.

\section{Summary of Results}

In this work, we have examined the general theory of second-order response functions. 
First, in Sec.~\ref{sec:2ndorder}, we exploited causality to introduce a spectral density representation of the second-order Kubo formula, generalizing the familiar Kramers-Kronig relations from linear response theory. 
The spectral representation Eq.~\eqref{eq:chi2specdensityrep} allowed us to express the large-frequency asymptotics of the second-order response function in terms of frequency moments of the spectral density. 
To relate these moments to experimentally-observable responses, we focused on second-harmonic generation. 
By using our explicit expression Eq.~\eqref{eq:chi2specdensity} for the spectral density to evaluate the frequency moments, we showed how they could be expressed in terms of averages of equal-time commutators. 
These relations are generalized sum rules relating equal-time averages, frequency moments of the spectral density, and large-frequency asymptotics of the second-harmonic generation response. 
We can summarize the sum rules for the second-order spectral density using its definition Eq.~\eqref{eq:chi2specdensity}
and repeated integration by parts via
\begin{align}\label{eq:chi2sumsummary}
\frac{1}{\pi^2}&\int d\alpha d\beta \left[\alpha^N\beta^M\chi^{(2)''}_{AB}(\alpha,\beta) \right] \nonumber\\
&= (i)^{N+M}\langle\com{\partial_t^M\com{\partial_t^NA}{B}}{B}\rangle_0.
\end{align}
These moments are related to the large-frequency asymptotic behavior of the second-order response via Eq.~\eqref{eq:chi2asymptotic} and \eqref{eq:shgasymptotic}. 
We give explicit expressions for the first three terms in the asymptotic expansion of the second harmonic generation response in Eqs.~\eqref{eq:shgnu0}--\eqref{eq:shgnu2}, our method in Sec.~\ref{sec:sumrule} straightforwardly extends to an infinite family of sum rules.

One should note the similarity between the sum rules in Eq.~\eqref{eq:chi2sumsummary} and the more familiar linear response sum rule
\begin{equation}\label{eq:linsumsummary}
-\frac{1}{\pi}\int  d\alpha \alpha^{N}\chi''_{AB}(\alpha) = i^N\langle\com{\partial_t^NA}{B}\rangle_0
\end{equation}
derived in Appendix~\ref{sec:linsum}. 
Both the linear sum rules Eq.~\eqref{eq:linsumsummary} and the second order sum rules Eq.~\eqref{eq:chi2sumsummary} control the asymptotic decay of response functions.

Next, we specialize in Sec.~\ref{sec:density} to the experimentally-relevant case of second-order density response. 
Applying our formalism to this case allowed us to derive nonlinear generalizations of the linear $f$-sum rule. 
In particular, we found in Eq.~\eqref{eq:shgrhonu1} that the leading-order behavior of the second-harmonic density response at large frequency is determined by the first moment of the spectral density; this can be expressed via our sum rule as the average second-order diamagnetic current (third derivative of the Hamiltonian with respect to vector potential). 
For nonrelativistic systems, this is identically zero and we have to go to next-order to find that our nonlinear sum rule relates the second moment of the spectral density response to the average charge density in equilibrium, in analogy with the linear $f$-sum rule. 
In particular, we found that the second-harmonic generation response for nonrelativistic systems satisfies
\begin{equation} 
\lim_{\omega\rightarrow\infty}\omega^4\bar{\chi}^{\mathrm{SHG}}_{\rho\rho}(\omega,\mathbf{q},\mathbf{q}) = \frac{3\bar{n}|\mathbf{q}|^4}{2m^2},
\end{equation}
which should be compared with the linear $f$-sum rule for nonrelativistic systems [see Eq.~\eqref{eq:fsumnr}]
\begin{equation}
\lim_{\omega\rightarrow\infty}\omega^2\chi_{\rho\rho}(\omega,\mathbf{q}) = |\mathbf{q}|^2\frac{\bar{n}}{m}.
\end{equation}
Thus, as for linear response, we find that the high-frequency behavior of the second harmonic generation response is universal for nonrelativistic systems and proportional to the ground state average density.

Next, we applied our results to a system of noninteracting electrons in a periodic potential in Sec.~\eqref{sec:freefermions}, showing that the leading-order sum rule can be expressed entirely in terms of matrix elements of the Bloch Hamiltonian. 
Using charge conservation, we showed how our second-order density response sum rules imply an infinite family of sum rules for the longitudinal second order conductivity, generalizing the results of Ref.~\cite{watanabe2020generalized} to nonzero wavevector and to higher moments of the conductivity.
Finally, we examined the rectification response, the zero frequency response to a harmonic perturbation at second order. 
We showed that for observables that can be written as time derivatives of other operators, the rectification response at large frequencies is determined by a set of sum rules expressible as averages of equal-time commutators.  

\section{Outlook}\label{sec:outlook}
Our work has implications for future theoretical and experimental studies on quantum materials. 
First, we re-emphasize that our spectral density and sum rules in Secs.~\ref{sec:2ndorder}, \ref{sec:sumrule}, \ref{sec:density}, \ref{sec:cond}, and \ref{sec:rect} are valid for any (energy-conserving) quantum system, and require no assumptions on the strength of electron-electron interactions or on the form of the unperturbed Hamiltonian. 
As such, our sum rules are formally exact, making them relevant to experiments on strongly-correlated as well as band materials. 
The second-order density response at large frequencies is directly accessible via nonlinear X-ray scattering, placing important constraints on such measurements, and enabling experimenters to calibrate the absolute scale of nonlinear cross sections in the same manner as linear spectroscopies \cite{Dressel2002,PaulMartin1968,Boothroyd2020}. 
 
 To emphasize the important role that sum rules play in interpreting spectroscopic data, recall that linear sum rules like the $f$-sum rule are essential for properly normalizing scattering data~\cite{shiles1980self,abbamonte2004imaging}. 
 Inelastic scattering intensity is, by Fermi's golden rule and the fluctuation dissipation theorem, proportional to the linear spectral density $\chi''_{\rho\rho}(\mathbf{q},\omega)$. 
 The proportionality constant is determined from microscopic details of the coupling to the spectroscopic probe, be it electrons in electron energy loss spectroscopy, neutrons in a neutron scattering experiment, or light in inelastic light scattering. 
 Because of this raw scattering intensity is given in unnormalized arbitrary units and cannot be compared across different experiments. 
 Since the $f$-sum rule gives a universal relation between the integrated structure factor to the ground state density, it can be used to normalize the data. 
 Even more, linear sum rules can be used as consistency relations that allow spectroscopic measurements in different frequency ranges obtained with different techniques to be stitched together to form composite datasets~\cite{smith1997dispersion}. 
 Our second order sum rules and Kramers-Kronig relations can provide experimentalists with similar tools for normalizing second harmonic generation responses and constructing broadband datasets of nonlinear optical properties of solids.

Additionally, Refs.~\cite{tam2022topological,tam2024topological} showed that $N$-point density-density correlation functions in a Fermi liquid are related to multipartite entanglement and Fermi surface topology in the ground state. 
Since these correlation functions appear in the Fourier transform of $\chi^{(2)''}_{\rho\rho}$ in the time-domain, the connection between these works and our sum rules is a fruitful area for future theoretical exploration.

Furthermore our general method in Secs.~\ref{sec:sumrule} and \ref{sec:rect} can be applied to derive sum rules for the conductivity tensor directly, rather than just its longitudinal components. 
For optical response at zero wavevector, our formalism of Sec.~\ref{sec:2ndorder} can be applied directly using $\mathbf{j}_{\mathbf{0}} = i[H_0,\mathbf{X}]$, where $\mathbf{X}$ is the many-body position operator, provided care is taken to also include the generalized diamagnetic conductivity (see, e.g.~the generalized current vertices in Refs.~\cite{parker2019diagrammatic,mckay2023spatially}). 
In this way our approach can bridge the gap between the Riemannian-geometric approach to nonlinear optical response from Ref.~\cite{ahn2022riemannian} and sum-rule based constraints generalizing the ``quantum weight'' of Refs.~\cite{souza2000polarization,onishi2023quantum,onishi2024quantum,verma2024instantaneous}.

Finally, let us remark on the applicability of the sum rules presented here to low-energy effective models, where very high-energy degrees of freedom have been projected out. 
In the context of such low-energy effective models, we typically consider the response of operators projected into the low-energy subspace (i.e.~projected density operators), and restrict frequency integrations in sum rules to the range of some gap or cutoff scale. 
For effective models whose low-energy subspace is spanned by ultralocalized Wannier functions, we expect our response functions and sum rules to hold in the form presented in Secs.~\ref{sec:sumrule} and \ref{sec:density}. 
This is because the algebra of projected density operators is unmodified provided all moments of the position operator in the Wannier basis are diagonal, as is the case in the ultralocalized (or tight-binding) limit~\cite{mckay2023spatially,bradlyn2022lecture,vanderbilt2018berry}. 
However, in moir\'{e} and other strongly-interacting topological systems, we are often interested in projecting onto low-energy degrees of freedom with delocalized Wannier functions, which is a situation in which low-energy partial sum rules are known to break down \cite{Meinders1993}. 
In such cases, a full theory of projected response functions is required. 
Preliminary work in this direction has focused on the density algebra in fractional quantum Hall~\cite{girvin1986magneto,haldane2011geometrical,gromov2017bimetric} and Chern insulators~\cite{roy2014band,neupert2012noncommutative,torma2022superconductivity}. 
Additionally, recent work on linear sum rules in Ref.~\cite{mendez2023theory} has shown the importance of the projected density operator on optical sum rules over restricted frequency range. 
Using the tools we developed here, the extension of these results to nonlinear response functions can be undertaken.

\begin{acknowledgments}
The authors thank C.~L. 
Kane, J.~E. 
Moore and P. 
Phillips for fruitful discussions. 
This work is supported by the U.S. 
DOE, Office of Basic Energy Sciences, Energy Frontier Research Center for Quantum Sensing and Quantum Materials through Grant No. 
DE-SC0021238. 
B.~B. received additional support for theoretical development from the Alfred P. 
Sloan foundation, and the National Science Foundation under grant DMR-1945058. 
P.~A. gratefully acknowledges additional support from the EPiQS program of the Gordon and Betty Moore Foundation, grant no. 
GBMF9452.
\end{acknowledgments}
\appendix

\section{Review of Linear Sum Rules}\label{sec:linreview}
In this Appendix, we review the derivation of the spectral density and general sum rules for linear response functions. 
We follow the approach of Refs.~\cite{kadanoff1963hydrodynamic,forster1975hydrodynamic}. 
Suppose our system of interest is governed by Hamiltonian $H_0$, and we apply a time-dependent external field that couples to an operator $B$ via the perturbing Hamiltonian
\begin{equation}
H_1 = f(t)B.
\end{equation}
We can then write the average of an operator $A$ to linear order in $B$ as
\begin{equation}\label{eq:chi1def}
\delta\langle A\rangle(t) = \langle A \rangle (t) -\langle A\rangle_0 = \int dt' \chi_{AB}(t-t') f(t'),
\end{equation} 
where $\langle A\rangle_0=\mathrm{tr}(A\rho_0)$ indicates the average of $A$ in the unperturbed equilibrium state $\rho_0$ and $\chi_{AB}(t-t')$ is known as the linear response function. 
For the remainder of this work we will assume that the equilibrium density matrix $\rho_0$ commutes with $H_0$, and so is time independent. 
The linear response function $\chi_{AB}(t-t')$ is given by the Kubo formula as
\begin{equation}\label{eq:kubo1timedomain}
\chi_{AB}(t-t') = -i\Theta(t-t')\langle\left[ A(t-t'),B(0)\right]\rangle_0,
\end{equation}
where $\Theta(t-t')$ is the Heaviside step function, and the time evolution of operators is evaluated in the Heisenberg picture with respect to the unperturbed Hamiltonian $H_0$. 
Assuming that $f(t)$ goes to zero infinitely far in the past (such that the system can be viewed as starting in the unperturbed ground state at $t\rightarrow-\infty$), we can use the convolution theorem to take the Fourier transform of Eq.~\eqref{eq:chi1def} to find
\begin{equation}
\delta\langle A\rangle(\omega) = \int dt e^{i\omega t} \delta\langle A\rangle(t) = \chi_{AB}(\omega) f(\omega),
\end{equation}
with 
\begin{align}\label{eq:chi1omegadef}
\chi_{AB}(\omega) &= \lim_{\epsilon\rightarrow 0} -i\int_0^\infty dt e^{i(\omega + i \epsilon)t}\langle\left[ A(t),B(0)\right]\rangle_0 \nonumber\\
&\equiv-i\int_0^\infty dt e^{i\omega^+t}\langle\left[ A(t),B(0)\right]\rangle_0.
\end{align}
In the second line of Eq.~\eqref{eq:chi1omegadef} we have introduced the shorthand $\omega^+ = \omega+i\epsilon$, where it is understood that the limit $\epsilon\rightarrow 0$ should be taken at the end of any calculation.

We now review some textbook properties of $\chi_{AB}(\omega)$, which can be found in, e.g. 
Refs.~\cite{kadanoff1963hydrodynamic,forster1975hydrodynamic,fetter2012quantum}. 
First, note that the Heaviside function in Eq.~\eqref{eq:kubo1timedomain} enforces causality: the response $\delta\langle A\rangle(t)$ depends only on the value of the perturbing field at earlier times $t' < t$. 
Writing
\begin{equation}\label{eq:chi1ift}
\chi_{AB}(t-t') =  \frac{1}{2\pi}\int d\omega \chi_{AB}(\omega) e^{-i\omega(t-t')},
\end{equation}
this implies that $\chi_{AB}(\omega)$ viewed as a function of complex $\omega_a+i\omega_b$ must be analytic in the upper half plane $\omega_b\geq 0$. 
To see this, we can imagine evaluating the inverse Fourier transform in Eq.~\eqref{eq:chi1ift} via contour integration. 
For $(t-t')<0$ we can close the contour in the upper half plane, since then the exponential factor is decaying for large $|\omega|$. 
This is guaranteed to give zero if there are no poles of $\chi_{AB}(\omega)$ in the upper half plane. 
We can verify this directly from Eq.~\eqref{eq:chi1omegadef} by inserting a complete set of eigenstates $\{\ket{n}\}$ of $H_0$ (with energies $E_n$) into the average. 
Taking the equilibrium density matrix to be given by Eq.~\eqref{eq:gndstate}, 
we have
\begin{widetext}
\begin{align}
\chi_{AB}(\omega) &=  -i\int_0^\infty e^{i\omega^+ t}\sum_{nm} p_n\left(\bra{n}A\ket{m}e^{i(E_n-E_m)t}\bra{m}B\ket{n} -  \bra{n}B\ket{m}\bra{m}A\ket{n}e^{i(E_m-E_n)t}\right) \\
&=\sum_{nm} p_n \left(\frac{\bra{n}A\ket{m}\bra{m}B\ket{n}}{\omega^+-(E_m-E_n)} - \frac{\bra{n}B\ket{m}\bra{m}A\ket{n}}{\omega^+-(E_n-E_m)}\right).\label{eq:chi1lehmann}
\end{align}
\end{widetext}
Eq.~\eqref{eq:chi1lehmann} is known as the spectral (or Lehmann) representation of $\chi_{AB}(\omega)$. 
We see explicitly that the poles of $\chi_{AB}$ in the complex $\omega$ plane occur when $\omega = E_n-E_m-i\epsilon$ for every pair of energies $E_n$ and $E_m$, and so are in the lower half plane as required by causality. 

Further exploiting the analytic properties of $\chi_{AB}(\omega)$ allows us to introduce a spectral density and derive Kramers-Kronig relations. 
We will take a somewhat unconventional path to these results. 
First, we derive an expression for $\Theta(t)e^{i\omega^+ t}$, which appears in the integrand of Eq.~\eqref{eq:chi1omegadef}:
\begin{equation}\label{eq:thetafnfourier}
\Theta(t)e^{i\omega^+ t} = \frac{1}{2\pi i}\int  d\alpha \frac{e^{i\alpha t}}{\alpha-\omega^+}.
\end{equation}
To verify this relation, we can evaluate the integral using contour integration in the complex $\alpha$ plane. 
When $t>0$, we can choose to close the integration contour in the upper half complex $\alpha$ plane (so that the exponential factor decays at infinity), and so pick up the residue at the pole $\alpha=\omega+i\epsilon$. 
On the other hand, when $t<0$, we must close the contour in the lower half plane, yielding zero since the integrand is analytic in the lower half plane. 
Using this relation, we can write
\begin{equation}\label{eq:specdensity1def}
\chi_{AB}(\omega) = \frac{1}{\pi}\int d\alpha \frac{\chi_{AB}''(\alpha)}{\alpha-\omega^+},
\end{equation}
where $\chi_{AB}''(\alpha)$ is known as the spectral density and given by 
\begin{equation}\label{eq:specdensity1formula}
\chi_{AB}''(\alpha) = -\frac{1}{2}\int dt e^{i\alpha t}\langle\left[ A(t),B(0)\right]\rangle_0.
\end{equation}
By inserting a complete set of states as we did in deriving the Lehmann representation Eq.~\eqref{eq:chi1lehmann} of $\chi_{AB}$, we can derive that the spectral density is given by
\begin{equation}\label{eq:specdensity1lehmann}
\chi_{AB}''(\alpha) = \pi\sum_{nm} \bra{n}A\ket{m}\bra{m}B\ket{n}\delta(\alpha+E_n-E_m) (p_m-p_n).
\end{equation}
We see that the spectral function is given by a weighted sum of $\delta$-functions. 
Note that when we consider response functions with $A=B$, then $\chi_{AA}''$ is real-valued. 
This is not true in general, however.

Next, we can use the Plemelj formula
\begin{equation}\label{eq:plemelj}
\frac{1}{\alpha-\omega^+} = \mathrm{P}\frac{1}{\alpha-\omega} + i\pi\delta(\alpha-\omega)
\end{equation}
(where $\mathrm{P}$ denotes the Cauchy principal value) to derive several useful properties of $\chi_{AB}$ and $\chi_{AB}''$. 
Inserting Eq.~\eqref{eq:plemelj} into Eq.~\eqref{eq:specdensity1def} we find
\begin{align}
\chi_{AB}(\omega) &= \frac{1}{\pi}\mathrm{P}\int d\alpha \frac{\chi_{AB}''(\alpha)}{\alpha-\omega} + i \chi_{AB}''(\omega) \label{eq:chi1intermsofspec}\\
&\equiv \chi_{AB}'(\omega) + i \chi_{AB}''(\omega),\label{eq:chiprimedef}
\end{align}
where we have defined $\chi_{AB}'(\omega)$ as the Hilbert transform of $\chi_{AB}''(\omega)$. 
When we consider response functions with $A=B$, we have from Eq.~\eqref{eq:chiprimedef} that $\chi_{AA}'(\omega)$ is the real part of $\chi_{AA}(\omega)$, and $\chi_{AA}''(\omega)$ is the imaginary part of $\chi_{AA}(\omega)$. 
For more general response functions, this does not hold.

By exploiting the analyticity of $\chi_{AB}(\omega)$ in the upper half plane, we also have that
\begin{equation}
0 = \frac{1}{2\pi i} \int d\alpha \frac{\chi_{AB}(\alpha)}{\alpha-\omega^-}.
\end{equation}
Using the Plemelj formula Eq.~\eqref{eq:plemelj} on the right hand side and comparing with Eqs.~\eqref{eq:chi1intermsofspec} and \eqref{eq:chiprimedef} yields the Kramers-Kronig relations
\begin{align}\label{eq:kk1relns}
\chi_{AB}'(\omega) &= \frac{1}{\pi}\mathrm{P}\int d\alpha \frac{\chi_{AB}''(\alpha)}{\alpha-\omega}, \\
\chi_{AB}''(\omega) &= -\frac{1}{\pi}\mathrm{P}\int d\alpha \frac{\chi_{AB}'(\alpha)}{\alpha-\omega}.
\end{align}

\subsection{Sum Rules}\label{sec:linsum}

Using Eq.~\eqref{eq:specdensity1def} we can derive an asymptotic expansion for $\chi_{AB}(\omega\rightarrow\infty)$ in terms of moments of the spectral density $\chi_{AB}''(\alpha)$. 
We can obtain an asymptotic expansion by taking $\omega\rightarrow\infty$ and Taylor expanding the denominator in Eq.~\eqref{eq:specdensity1def}. 
We find
\begin{align}\label{eq:linearasymptotic}
\chi_{AB}(\omega\rightarrow\infty)&\sim \frac{1}{\pi}\int d\alpha\sum_{n=0}^{\infty} \frac{-\alpha^n}{\omega^{n+1}}\chi_{AB}''(\alpha) \nonumber\\
&\sim \sum_{n=0}^{\infty} \frac{1}{\omega^{n+1}}\left[-\frac{1}{\pi}\int d\alpha \alpha^n\chi_{AB}''(\alpha)\right]\nonumber \\
&\sim \sum_{n=0}^{\infty} \frac{\mu_{AB}^{(n)}}{\omega^{n+1}}.
\end{align}
From Eq.~\eqref{eq:linearasymptotic} we see that the $n$-th moments $\mu_{AB}^{(n)}$ of $\chi_{AB}''$ determine the large-frequency asymptotic decay of the response function $\chi_{AB}(\omega\rightarrow\infty)$. 
Furthermore, since $\chi''_{AB}(\alpha)$ is given in Eq.~\eqref{eq:specdensity1formula} as the Fourier transform of a correlation function, we can integrate by parts to express $\mu_{AB}^{(n)}$ as an equal time average
\begin{align}\label{eq:linsumrule}
\mu_{AB}^{(n)} &= -\frac{1}{\pi}\int  d\alpha \alpha^{n}\chi''_{AB}(\alpha) \nonumber\\
&=\frac{1}{2\pi}\int  d\alpha   dt \alpha^{n} e^{i\alpha t}\langle\left[ A(t),B(0)\right]\rangle_0\nonumber \\
&=\langle\left[(i\partial_t)^n A(0),B(0)\right]\rangle_0\nonumber \\
&=(-1)^n\langle\left[\underbrace{\left[H,\left[H,\dots \left[H,A\right]\right]\right]}_{\text{$n$ times}},B\right]\rangle_0.
\end{align}
Eq.~\eqref{eq:linsumrule} gives a family of sum rules, relating the $n$-th moment of the spectral function to the equal-time average of the commutator of the $n$-th time derivative of $A$ with $B$. 
Hence the sum rule also expresses the asymptotic decay of the response function $\chi_{AB}$ at large $\omega$ to the same equal time commutator, via Eq.~\eqref{eq:linearasymptotic}. 
Importantly, Eqs.~\eqref{eq:linsumrule} and \eqref{eq:linearasymptotic} are \emph{exact} relations valid for any quantum system and any response function. 
They hold for both noninteracting and strongly interacting electron systems, and represent one of the few exact constraints on measurable quantities.

\subsection{Application: Density-Density Response}\label{sec:rhorho1}

As an illustrative example, let us consider the density-density response function
\begin{equation}
\chi_{\rho\rho}(\mathbf{q},\omega) \equiv \chi_{\rho_\mathbf{q}\rho_{-\mathbf{q}}}(\omega)
\end{equation}
that governs the response of the Fourier component $\rho_{\mathbf{q}}$ of the average density with wavevector $\mathbf{q}$ to perturbations that couple to the particle density. 
The corresponding spectral density is given by
\begin{equation}
\chi_{\rho\rho}''(\mathbf{q},\alpha) = -\frac{1}{2}\int dt e^{i\alpha t}\langle\left[\rho_{\mathbf{q}}(t),\rho_{-\mathbf{q}}(0)\right]\rangle_0.
\end{equation}
We can now evaluate the moments $\mu_{\rho\rho}^{(n)}(\mathbf{q})$ of the spectral function $\chi_{\rho\rho}''$. 
We will focus on the lowest two cases, $n=0$ and $n=1$. 
For $n=0$ we find
\begin{align}\label{eq:mu0rho}
\mu_{\rho\rho}^{(0)}(\mathbf{q}) &= -\frac{1}{\pi}\int  d\alpha \chi_{\rho\rho}''(\mathbf{q},\alpha)\nonumber \\
&= \langle\left[\rho_{\mathbf{q}},\rho_{-\mathbf{q}}\right]\rangle_0 \nonumber\\
&= 0,
\end{align}
where we have used the fact that the density operators commute with each other at equal time, since they are functions only of the position operator (note that this relation fails to hold for the density operator projected into a set of low-energy bands). 
Turning next to $n=1$, we have
\begin{align}\label{eq:mu1rho}
\mu_{\rho\rho}^{(1)}(\mathbf{q}) &= -\frac{1}{\pi}\int  d\alpha \alpha\chi_{\rho\rho}''(\mathbf{q},\alpha)\nonumber \\
&= \langle\left[i\partial_t\rho_{\mathbf{q}},\rho_{-\mathbf{q}}\right]\rangle_0.
\end{align}
To simplify this further, we recall that charge conservation implies
\begin{equation}\label{eq:continuity}
i\partial_t\rho_{\mathbf{q}} = -[H,\rho_\mathbf{q}] = \mathbf{q}\cdot\mathbf{j}_{\mathbf{q}},
\end{equation}
where $\mathbf{j}_\mathbf{q}$ is the $\mathbf{q}$-th Fourier component of the current density operator. 
Inserting this into Eq.~\eqref{eq:mu1rho} we find
\begin{equation}\label{eq:fsum1general}
\mu_{\rho\rho}^{(1)}(\mathbf{q}) = \langle\left[\mathbf{q}\cdot\mathbf{j}_\mathbf{q},\rho_{-\mathbf{q}}\right]\rangle_0.                        
\end{equation}
Eq.~\eqref{eq:fsum1general} is the general form of the $f$-sum rule. 
For nonrelativistic (n.r.) systems where the current density is proportional to the momentum density, the canonical commutation relations allow us to write the right hand side as 
\begin{equation}\label{eq:fsumnr}
\mu_{\rho\rho}^{(1)}(\mathbf{q})\rightarrow_{\text{n.r.}} \frac{|\mathbf{q}|^2\bar{n}}{m},
\end{equation}
where $m$ is the particle mass and $\bar{n}$ is the average ground state density $\bar{n} = \langle \rho_\mathbf{q=0} \rangle_0$.

More generally, we can follow Ref.~\cite{mckay2023spatially} to relate the average in Eq.~\eqref{eq:fsum1general} to the diamagnetic current. 
Concretely, note that for any operator $\mathcal{O}(\{\hat{\mathbf{x}}_i\},\{\hat{\mathbf{p}_i}\})$ 
that depends on the position and momentum operators for each particle in the system (indexed by $i$), we can define an operator $\mathcal{O}_A = \mathcal{O}(\{\hat{\mathbf{x}}_i\},\{\hat{\mathbf{p}}_i-\mathbf{A}(\mathbf{\hat{x}}_i)\})$ 
that is minimally coupled to a background electromagnetic vector potential $\mathbf{A}$. 
Writing the density operator as
\begin{equation}
\rho_\mathbf{q} = \frac{1}{V}\sum_i e^{-i\mathbf{q}\cdot\mathbf{x}_i},
\end{equation}
where $V$ is the volume of the system, we have using the canonical commutation relations that
\begin{equation}\label{eq:gaugevariation}
\left[\mathcal{O},\rho_\mathbf{q}\right] = \left. \frac{\delta \mathcal{O}_A}{\delta A^\mu_{-\mathbf{q}}}\right|_{\mathbf{A}\rightarrow 0}q_\mu.
\end{equation}
Applying Eq.~\eqref{eq:gaugevariation} to both the continuity equation Eq.~\eqref{eq:continuity} and to the sum rule Eq.~\eqref{eq:fsum1general} we find
\begin{equation}\label{eq:fsumdiamagnetic}
\mu_{\rho\rho}^{(1)}(\mathbf{q}) = q_\mu q_\nu\langle\left.\frac{\delta^2 H_A}{\delta A^\mu_\mathbf{q} A^\nu_{\mathbf{-q}}}\right|_{\mathbf{A}=0}\rangle_0,
\end{equation}
where $H_A$ is the Hamiltonian $H_0$ minimally coupled to the electromagnetic vector potential $\mathbf{A}$. 
Eq.~\eqref{eq:fsumdiamagnetic} shows that the first moment $\mu_{\rho\rho}^{(1)}(\mathbf{q})$ of the spectral density $\chi_{\rho\rho}''(\mathbf{q},\omega)$ is proportional to the average of the diamagnetic current 
\begin{equation}
\left.\frac{\delta^2 H_A}{\delta A^\mu_\mathbf{q} A^\nu_{\mathbf{-q}}}\right|_{\mathbf{A}=0}.
\end{equation}
Combining Eqs.~\eqref{eq:fsumdiamagnetic}, \eqref{eq:mu0rho}, \eqref{eq:mu1rho}, and \eqref{eq:linearasymptotic}, we have that the high-frequency asymptotic expansion of the density-density response function is given by
\begin{equation}\label{eq:fsumasymptotic}
\chi_{\rho\rho}(\mathbf{q},\omega\rightarrow\infty) \sim \frac{q_\mu q_\nu}{\omega^2}\langle\left.\frac{\delta^2 H_A}{\delta A^\mu_\mathbf{q} A^\nu_{\mathbf{-q}}}\right|_{\mathbf{A}=0}\rangle_0.
\end{equation}

\section{Derivation of the Second-Order Kubo Formula}\label{sec:derivations}

In this section, we derive expressions for the linear and second order response functions for the average of an operator $A$ in response to a perturbation
\begin{equation}
H_1(t) = g(t)B.
\end{equation}
We will assume for now that the form of the operator $A$ does not depend of $g(t)$, so that we do not have any ``contact'' (diamagnetic) terms. 
We assume that at large negative times $t\rightarrow -\infty$ that the system is in a steady state of the unperturbed Hamiltonian $H_0$ with density matrix $\rho_0$ such that $[H_0,\rho_0]=0$. 
We also assume that $g(t)$ goes to zero as $t\rightarrow -\infty$, such that we can write 
\begin{equation}\label{eq:adiabaticdef}
g(t) = e^{\epsilon t}f(t)
\end{equation}
with $\epsilon$ a positive infinitesimal that is taken to zero at the end of any calculation. 
Eq.~\eqref{eq:adiabaticdef} ensures that the perturbation $f(t)$ turns on adiabatically. 

We define the linear and second order response functions $\chi_{AB}(t-t')$ and $\chi^{(2)}_{AB}(t-t',t-t'')$ via a series expansion of $\langle A\rangle(t)$ in powers of $f(t)$. 
Concretely,
\begin{widetext}
\begin{equation}\label{eq:responsefnsgeneraldef}
\langle A\rangle(t) = \langle A\rangle_0 + \int  dt' \chi_{AB}(t-t')f(t') + \int  dt'\int dt'' \chi^{(2)}_{AB}(t-t',t-t'')f(t')f(t'')+\dots.
\end{equation}
\end{widetext}
To evaluate the response functions, we can solve the equations of motion for the density matrix $\rho(t)$ characterizing the state of the system in the presence of the perturbation $H_1$, subject to the initial conditions $\rho(t\rightarrow-\infty)=\rho_0$. 
Writing
\begin{equation}\label{eq:rhoexpansion}
\rho(t) = \rho_0 + \rho_1(t) +\rho_2(t)+\dots,
\end{equation}
where the subscript indicates the order of $f$ on which each term depends, we can write the equations of motion for $\rho(t)$ as
\begin{align}
\frac{\partial}{\partial t} &[\rho_0 + \rho_1(t) +\rho_2(t)+\dots] = -i[H,\rho] \nonumber\\
&=-i[H_0,\rho_1] -i[H_1,\rho_1] - i[H_0,\rho_2]+\dots.
\end{align}
Equating powers of $f(t)$, we arrive at the following equations for $\rho_1(t)$ and $\rho_2(t)$:
\begin{align}
    \frac{\partial \rho_1}{\partial t}&=-i\com{H_0}{\rho_1}-i\com{H_1}{\rho_0},\label{eq:rho1eq} \\
    \frac{\partial \rho_2}{\partial t}&=-i\com{H_0}{\rho_2}-i\com{H_1}{\rho_1}.\label{eq:rho2eq}
\end{align}
We can first solve Eq.~\eqref{eq:rho1eq} by switching to the interaction picture, following Ref.~\cite{fetter2012quantum}. 
We find
\begin{align}
\rho_1(t) &= -i\int_{-\infty}^t dt' \com{e^{i(t'-t)H_0}H_1(t')e^{-i(t'-t)H_0}}{\rho_0}\nonumber \\
&=-i\int_{-\infty}^t dt'\com{B(t'-t)}{\rho_0}f(t')e^{\epsilon t'},\label{eq:rho1solved}
\end{align}
where the time dependence in Eq.~\ref{eq:rho1solved} is evaluated using the unperturbed Hamiltonian $H_0$. 
Inserting Eq.~\eqref{eq:rho1solved} into Eq.~\eqref{eq:rho2eq}, we can solve for $\rho_2(t)$ in the same way. 
We find that
\begin{widetext}
\begin{align}
\rho_2(t) &= -i\int_{-\infty}^t dt' e^{-iH_0t}\com{e^{it'H_0}H_1(t')e^{-it'H_0}}{e^{iH_0t'}\rho_1(t')e^{-iH_0t'}}e^{iH_0t}\nonumber \\
&= -\int_{-\infty}^t dt'\int_{-\infty}^{t'} dt'' \com{B(t'-t)}{\com{B(t''-t)}{\rho_0}}f(t')f(t'')e^{\epsilon(t'+t'')}.
\end{align}

Using Eq.~\eqref{eq:rhoexpansion} to write the average $\langle A\rangle(t)$ as
\begin{equation}
\langle A\rangle(t) = \langle A\rangle_0 + \mathrm{tr}(A\rho_1(t)) +\mathrm{tr}(A\rho_2(t))+\dots,
\end{equation}
we can equate terms with Eq.~\eqref{eq:responsefnsgeneraldef} order by order in $f(t)$ to extract the response functions. 
For the linear response function
\begin{align}\label{eq:derived linearresp}
\int  dt' \chi_{AB}(t-t')f(t') &= \mathrm{tr}(A\rho_1(t)) \nonumber\\
&= -i\int_{-\infty}^t dt'\mathrm{tr}\left(A\com{B(t'-t)}{\rho_0}\right)f(t')e^{\epsilon t'} \nonumber\\
&=-i\int_{-\infty}^t dt'\langle\com{A(0)}{B(t'-t)}\rangle_0 f(t')e^{\epsilon t'} \nonumber\\
&=-ie^{\epsilon t}\int dt' \Theta(t-t')\langle\com{A(t)}{B(t')}\rangle_0 f(t')e^{-\epsilon (t-t')},
\end{align}
where we have used time-translation invariance to shift the time arguments within the average by $t$. 
Eq.~\eqref{eq:derived linearresp} yields immediately the Kubo formula Eq.~\eqref{eq:kubo1timedomain}.
For the second-order response, we have
\begin{align}\label{eq:derived nonlinearresp}
\int  &dt' dt'' \chi^{(2)}_{AB}(t-t',t-t'')f(t')f(t'') =\mathrm{tr}(A\rho_2(t)) \nonumber\\
&=  -\int_{-\infty}^t dt'\int_{-\infty}^{t'} dt'' \mathrm{tr}(A\com{B(t'-t)}{\com{B(t''-t)}{\rho_0}})f(t')f(t'')e^{\epsilon(t'+t'')}\nonumber \\
&=-e^{2\epsilon t}\int dt'  dt''\Theta(t-t')\Theta(t'-t'')\langle\com{\com{A(t)}{B(t')}}{B(t'')}\rangle_0e^{-\epsilon(2t-t'-t'')}f(t')f(t'').
\end{align}
We thus find that
\begin{equation}\label{eq:chi2timedomain}
\chi^{(2)}_{AB}(t-t',t-t'') = -\Theta(t-t')\Theta(t'-t'')\langle\com{\com{A(t)}{B(t')}}{B(t'')}\rangle_0e^{-\epsilon(2t-t'-t'')}.
\end{equation}
As in linear response, it will be useful to Fourier transform Eq.~\eqref{eq:derived nonlinearresp} to arrive at an expression for the frequency-dependent nonlinear response. 
Concretely, we find
\begin{align}
\frac{1}{2\pi}\int  dt e^{i\omega t}\mathrm{tr}(\rho_2(t)A)e^{-2\epsilon t} &=\frac{1}{2\pi}\int  dt   dt' dt'' \chi^{(2)}_{AB}(t-t',t-t'')f(t')f(t'')e^{i\omega t}\nonumber \\
&=\frac{1}{2\pi}\int  dt  dt' dt''  d\omega_1  d\omega_2 \chi^{(2)}_{AB}(t-t',t-t'')f(\omega_1)f(\omega_2)e^{i\omega t}e^{-i\omega_1 t'} e^{-i\omega_2t''}\nonumber \\
&=\int  d\omega_1 d\omega_2 \delta(\omega-\omega_1-\omega_2)\chi^{(2)}_{AB}(\omega_1,\omega_2)f(\omega_1)f(\omega_2),
\end{align}
where we have defined
\begin{align}
\chi^{(2)}_{AB}(\omega_1,\omega_2) &= \int dx\int dy \chi^{(2)}_{AB}(t_1,t_2) e^{i\omega_1 t_1}e^{i\omega_2 t_2}\nonumber \\
&=-\int dt_1\int dt_2 \Theta(t_1)\Theta(t_2-t_1)\langle\com{\com{A(t_2)}{B(t_2-t_1)}}{B(0)}\rangle_0e^{i\omega_1^+ t_1}e^{i\omega_2^+ t_2}.\label{eq:primitivesecondorder}
\end{align}
Note that because the operators $A$ and $B$ do not necessarily commute with themselves nor each other at different times, we cannot extend the domain of time integration by symmetrizing under $\omega_1\leftrightarrow\omega_2$ (in contrast to the case of time-ordered correlation functions).

To make the causal structure of $\chi^{(2)}_{AB}(\omega_1,\omega_2)$ manifest, it is useful to perform an additional change of variables
\begin{align}
u&=t_1,\nonumber \\
v&=t_2-t_1
\end{align}
in Eq.~\eqref{eq:primitivesecondorder} to find
\begin{equation}\label{eq:secondorderw1w2}
\chi^{(2)}_{AB}(\omega_1,\omega_2)=-\int_{0}^{\infty}du\int_{0}^{\infty}dv \langle\com{\com{A(u+v)}{B(v)}}{B(0)}\rangle_0e^{i\omega_{12}^+ u}e^{i\omega_2^+ v},
\end{equation}
which is the form of $\chi^{(2)}_{AB}(\omega_1,\omega_2)$ introduced in the main text Eq.~\eqref{eq:chi2finalmain}.
\end{widetext}
To conclude, let us note that causality places analyticity constraints on $\chi^{(2)}_{AB}(\omega_1,\omega_2)$. 
First, note from Eq.~\eqref{eq:chi2timedomain} that $\chi^{(2)}_{AB}(t_1,t_2)$ vanishes for $t_1<0$ and for $t_2<0$ (in fact, it vanishes whenever $t_2<t_1$). 
Writing
\begin{equation}
\chi^{(2)}_{AB}(t_1,t_2) = \frac{1}{4\pi^2}\int d\omega_1\int d\omega_2 \chi^{(2)}_{AB}(\omega_1,\omega_2)e^{-i(\omega_1 t_1+\omega_2t_2)}
\end{equation}
and evaluating the integrals by contour integration, we deduce that $\chi^{(2)}_{AB}(\omega_1,\omega_2)$ is analytic when both $\mathrm{Im}(\omega_1)>0$ and $\mathrm{Im}(\omega_2)>0$.

\bibliography{refs.bib}

\end{document}